\documentclass[aps,
		prd,
		reprint,
        superscriptaddress,
        shortbibliography,
		  nofootinbib,
		floatfix,
		notitlepage, onecolumn
		]{revtex4-2}

\usepackage{amsmath}
\usepackage{mathtools}
\usepackage{amssymb}
\usepackage{mathrsfs}
\usepackage{slashed}
\usepackage{color}
\usepackage{xcolor}
\usepackage{hyperref}
\usepackage{braket}
\hypersetup{colorlinks=true,linkcolor=blue} 
\usepackage{comment}
\usepackage{bm}
\usepackage[normalem]{ulem}
\usepackage{bbold}

\newcommand{\mathbfbs}[1]{\boldsymbol{\mathbf{#1}}}
\newcommand{\radpr}{%
\hspace{.2ex}\begin{tikzpicture}%
\draw[draw=blue, smallsnake] (-.5ex,0) -- (.5ex,0);%
\draw[double] (0,.5ex) arc (90:270:.5ex);%
\draw[double] (0,.5ex) -- (.5ex,.5ex);
\draw[double] (0,-.5ex) -- (.5ex,-.5ex);
\end{tikzpicture}%
}
\newcommand{\pair}{%
\hspace{.2ex}\begin{tikzpicture}%
\draw[double] (0,.5ex) arc (90:270:.5ex);%
\draw[double] (0,.5ex) -- (.5ex,.5ex);
\draw[double] (0,-.5ex) -- (.5ex,-.5ex);
\end{tikzpicture}%
}
\newcommand{\pairsqr}{%
\hspace{.2ex}\begin{tikzpicture}%
\draw[fill=gray, draw=gray] (0,0) circle (.2ex);%
\draw (0,0) -- (.75ex,.75ex);
\draw (0,0) -- (.75ex,.25ex);
\draw (0,0) -- (.75ex,-.25ex);
\draw (0,0) -- (.75ex,-.75ex);
\end{tikzpicture}%
}
\newcommand{\ccd}{%
\hspace{.2ex}\begin{tikzpicture}%
\draw[double] (0,0) circle (.4ex);%
\end{tikzpicture}%
}
\newcommand{\tadpl}{%
\hspace{.2ex}\begin{tikzpicture}%
\draw[draw=blue, smallsnake] (.25ex,0) -- (1.25ex,0);%
\draw[double] (0,0) circle (.25ex);
\end{tikzpicture}%
}
\newcommand{\forwarde}{%
\hspace{.1ex}
\begin{tikzpicture}%
\draw (0,.077ex) -- (1.2ex,.077ex);
\draw (0,-.077ex) -- (1.2ex,-.077ex);
\draw (.45ex,.33ex)--(.7ex,0);
\draw (.45ex,-.33ex)--(.7ex,0);
\end{tikzpicture}%
}
\newcommand{\forwardp}{%
\hspace{.1ex}
\begin{tikzpicture}%
\draw (0,.077ex) -- (1.2ex,.077ex);
\draw (0,-.077ex) -- (1.2ex,-.077ex);
\draw (.5ex,0)--(.75ex,.33ex);
\draw (.5ex,0)--(.75ex,-.33ex);
\end{tikzpicture}%
}

\newcommand{\be}{\begin{equation}}
\newcommand{\ee}{\end{equation}}
\newcommand{\bi}{\begin{enumerate}}
\newcommand{\ei}{\end{enumerate}}
\newcommand{\ud}{{\mathrm{d}}}
\hfuzz=\maxdimen \tolerance=10000
\vfuzz=\maxdimen \tolerance=10000
\newcommand{\Apair}[2]{\mathcal{A}_{\pair}(#1,#2)}
\newcommand{\boldpair}{\mathbfbs{\mathcal{A}}_{\pair}}

\usepackage{feynmp-auto}
\usepackage{tikz}
\usetikzlibrary{decorations.pathmorphing,decorations.markings,arrows}
\tikzset{snake it/.style={decorate, decoration={snake,segment length=4}}}
\tikzset{smallsnake/.style={decorate, decoration={snake,segment length=2, amplitude=1}}}

\begin{document}

\title{Pair creation, backreaction and resummation in strong fields}

\author{Patrick Copinger}
\email{patrick.copinger@plymouth.ac.uk}
\affiliation{Centre for Mathematical Sciences, University of Plymouth, Plymouth, PL4 8AA, UK}
\author{James P. Edwards}
\email{james.p.edwards@plymouth.ac.uk}
\affiliation{Centre for Mathematical Sciences, University of Plymouth, Plymouth, PL4 8AA, UK}
\author{Anton Ilderton}
\email{anton.ilderton@ed.ac.uk}
\affiliation{Higgs Centre, School of Physics and Astronomy, University of Edinburgh, EH9 3FD, UK}
\author{Karthik Rajeev}
\email{karthik.rajeev@ed.ac.uk}
\affiliation{Higgs Centre, School of Physics and Astronomy, University of Edinburgh, EH9 3FD, UK}

\begin{abstract}
We revisit particle creation in strong fields, and backreaction on those fields, from an amplitudes perspective. We describe the strong field by an initial coherent state of photons which we explicitly evolve in time, thus going beyond the background field approximation, and then consider observables which quantify the effects of backreaction. We present expressions for the waveform, vacuum persistence probability and number of produced photons at next-to-leading-order, all of which are impacted by backreaction, along with the number and statistics of produced pairs. We find that converting between in-out (amplitude) and in-in (expectation value) expressions requires explicit resummation of an infinite number of disconnected loop diagrams.
\end{abstract}

\maketitle
\onecolumngrid

\section{Introduction}

The intense electromagnetic fields of modern lasers provide opportunities for studying QED in the largely untested strong-field regime~\cite{Cole:2017zca,Poder:2017dpw,Abramowicz:2021zja,Ahmadiniaz:2024xob,Los:2024ysw}. The majority of theoretical approaches to `strong field QED' (SFQED) use the Furry expansion~\cite{Furry:1951bef}, in which the strong field is represented as a fixed, classical background, about which field perturbations are assumed small. For reviews see~\cite{Gonoskov:2021hwf,Fedotov:2022ely}. This is a good approximation in a broad range of scenarios. However, backreaction on the field, in the form of e.g.~beam-depletion, is expected to become important in collisions of high-charge bunches with very intense laser fields~\cite{Seipt:2016fyu}, and in high-multiplicity cascades where significant energy is taken from the driving fields~\cite{BellKirk:2008,Fedotov:2010ja,Bulanov:2013cga,Mironov:2021fft,Cruz:2022zyp,Mercuri-Baron:2024rdc}. It may also be necessary to account for backreaction in the high-intensity regime where the loop expansion of SFQED must be resummed to extract sensible physical predictions~\cite{Ritus1,Narozhnyi:1980dc,Fedotov:2016afw}. Going beyond the background field approximation, and treating depletion analytically is, however, challenging~\cite{
Seipt:2016fyu,
Luo:2018rkp}.
There have been some attempts to investigate how a pair-creating field depletes in scalar QED -- see \cite{PhysRevD.40.456}, for QED$_{2}$ see~\cite{QED2,QED22}, and see~\cite{QCD} for QCD. 

Here we approach the topic of backreaction slowly but systematically, staying as close as possible to standard methods of SFQED. We take the strong field to be initially described by a coherent state of
photons.
We evolve this state in time, and calculate observables as expectation values in the time-evolved state: we emphasise that we make no assumption on how the coherent state evolves, whereas in the standard approach to SFQED one effectively assumes it remains unchanged, which is equivalent to assuming a fixed background~\cite{Frantz,Kibble:1965zza,Gavrilov:1990qa}. As such we will be able to observe changes to the coherent state, i.e.~backreaction on the initial strong field. Our focus will be on initial coherent states which can spontaneously create particle/anti-particle pairs (via the Schwinger effect), hence state evolution will clearly be non-trivial. A coherent state approach to  depletion has been adopted in the past: in~\cite{Ilderton:2017xbj} depletion in laser-matter interactions was modelled in an SFQED context as a deformation of the initial coherent state; and in \cite{Dvali1, Dvali2} backreaction was associated to corrections in the inverse of the expected coherent state occupation number, $N$ about the large $N$ limit of scattering processes. See~\cite{Ekman:2020vsc} for an exactly solvable toy model of backreaction using coherent states.

In order to perform these calculations we need an expansion scheme, which we choose so as to parallel the Furry expansion (in which, concretely, the background is fixed but treated exactly, while quantised photons are treated with the familiar perturbative expansion in the fine structure constant)~\cite{Furry:1951bef}. This approach will have the benefit of allowing us to match to familiar objects, while seeing explicitly what the background field approximation misses.

We will take an amplitudes-based approach to this material, which will expose a great deal of structure, and apply our methods to observables that are not often considered in SFQED.
One such observable, and one which naturally exhibits effects due to backreaction, is the \emph{waveform}, the leading asymptotic form of (the expectation value of) the electromagnetic field $F_{\mu\nu}$ in the system. The corresponding gravitational waveform is commonly studied in the context of the classical two-body problem~\cite{Cristofoli:2021vyo, Bautista:2021inx,Adamo:2022qci}. Waveforms generated by purely quantum effects, on the other hand, are less commonly explored.

In terms of structure, we will see that, due to the possibility of particle creation, observables pick up contributions from \emph{non-trivial} bubble diagrams which mix the loop and Furry expansions. This implies non-trivial disconnected contributions to observables, which are also not often considered in SFQED due to the common use of plane-wave fields, which cannot produce pairs~\cite{Schwinger:1951nm}. Furthermore, this investigation will provide us with an accessible model in which to explore resummation. This is a topic which has arisen in many SFQED contexts in recent years, including the infra-red~\cite{Ilderton:2012qe}, effective actions~\cite{Karbstein:2019wmj}, bubble chains~\cite{Mironov:2020gbi,Mironov:2021ohk}, collinear contributions~\cite{Edwards:2020npu}, resurgence~\cite{Dunne:2021acr,Dunne:2022esi}, radiation reaction~\cite{Torgrimsson:2021wcj,Torgrimsson:2021zob,Torgrimsson:2022ndq}, and particle states~\cite{Podszus:2021lms,Podszus:2022jia}.

This paper is organised as follows. In Sec.~\ref{sec:review} we review the calculation of scattering amplitudes involving coherent states, and show how resulting observables may be related to familiar background-field amplitudes, and calculated order by order in analogy to the Furry expansion. In subsequent sections we explore various observables in this expansion. Working first to leading order in Sec~\ref{sec:leading}, we discuss pair production and relate the amplitudes-based approach to that based on Bolgoliubov transformations. We proceed to next-to leading order (NLO) in Sec~\ref{sec:nlo}, where backreaction on the driving fields first appears. We consider first the waveform, in which interference effects signal deviations from the background-field approximation. To give a completely explicit example, we calculate the waveform explicitly in a simple pair-producing field model. In Sec.~\ref{sec:resum} we re-analyse several of our results in the context of resummation, asking how our results are built from  Feynman diagrams. We will see that even well-known results such as the vacuum persistence amplitude and the number of created pairs exhibit an extremely rich structure in terms of disconnected and cut diagrams. We conclude in Sec.~\ref{sec:concs}. For simplicity, we restrict to scalar QED throughout, and we set $\hbar=c=1$.

\section{Setup and review}\label{sec:review}

The physical effects of interest in this paper arise in the presence of strong electromagnetic fields. To model these fields we take the initial scattering state to be a \emph{coherent state} of photons~\cite{sudarshan:1963es,glauber:1963ci,Frantz,Kibble:1965zza}. Being of minimal uncertainty and arbitrarily high occupancy~\cite{GerryKnight2004}, these states are a natural choice to describe classical fields. We begin by briefly reviewing the necessary properties of coherent states.

\subsection{The initial coherent state}
In terms of the mode operator $a_s(k)$ and polarisation $\varepsilon^s_\mu(k)$ for photons of momentum $k$ and helicity $s$, the mode expansion of the free electromagnetic field is
\begin{align}\label{A-expansion}
    A_{\mu}(x) = 
    \int_k
    \big [a_{s}(k)\varepsilon^{s}_{\mu}(k)e^{-ik\cdot x}+a^{\dagger}_{s}(k){\bar\varepsilon}^{s}_{\mu}(k)e^{i k\cdot x}\big] \;,
\end{align}
where an integral with subscript denotes the usual on-shell momenta integral,
\be
    \int_k \equiv \int \!\frac{\ud^3{\mathbf{k}}}{(2\pi)^3 2k_0} \;,
\ee
with $k_0$ the on-shell energy (for photons or, below, massive fields as appropriate). To avoid the appearance of many sums, we adopt the convention that if the helicity label ($s$, $\sigma$, etc) appears twice in any expression then it is summed over, \emph{whether it appears as an index or otherwise}. The mode operators obey $[a_s(k),a^\dagger_{\sigma}(k')] = \delta_{s\sigma}{\delta}_{k'k}$, in which $\delta_{k'k}$ should be understood as the on-shell delta function, i.e.$~\int_k \delta_{k'k} = 1$. A coherent state of photons may then be written~\cite{glauber:1963ci,GerryKnight2004}
\be
    \ket{z} \equiv \mathcal{D}(z) \ket{0} \;, \qquad
    \mathcal{D}(z) = \exp\bigg[
    \int_k  a^\dagger_{s}(k)z_s(k)
    -
    a_{s}(k){\bar z}_{s}(k)
    \bigg] \;,
\ee
in which $z_s(k)$ are two complex profile functions.
Coherent states are not eigenstates of number operators $N(k) = a_{s}^{\dagger}(k) a_{s}(k)$; rather, exponentiation means that they are made up of states states with arbitrary high occupation numbers. Taking the expectation value of (\ref{A-expansion}) in the coherent state $|z\rangle$ simply replaces the mode operators by the profile functions, $z_{s}(k)$, yielding a classical vacuum solution of Maxwell's equations associated to this choice of profile:
\be\label{Acl-def}
\bra{z}A_\mu(x) \ket{z} =
    \int_k 
    \big [z_{s}(k)\varepsilon^{s}_{\mu}(k)e^{-ik\cdot x}+{\bar z}_{s}(k){\bar\varepsilon}^{s}_{\mu}(k)e^{i k\cdot x}\big]
     =:   A^\text{cl}_\mu(x) \;.  
\ee
For example, the choice $z_s(k)\sim \delta(k_1)\delta(k_2)f(k_3)$ will lead to a plane wave $A_\mu^\text{cl}$ with pulse profile determined by $f(k_3)$, while less singular choices of $z_s(k)$ lead to spatially focussed fields, see e.g.~\cite{Fedotov2009exact}.
The essential property of the `displacement operator' $\mathcal{D}(z)$ required for all calculations with coherent states is that it acts on mode operators as~\cite{GerryKnight2004}
\begin{align}
    {\mathcal{D}}^{\dagger}[z] a_s(k) \mathcal{D}[z]
    &=
    a_{s}(k)+z_s(k) \;.
\end{align}
This in turn implies that for an arbitrary functional $\mathcal{F}$ of the gauge field, we have the relation
\begin{align}\label{A-shift}
    {\mathcal{D}}^{\dagger}[z]\mathcal{F}[A]\mathcal{D}[z]
    =
    \mathcal{F}[A+A^{\text{cl}}] \;.
\end{align}
This shift of the dependence on a quantised field by a classical field configuration can be interpreted as introducing a background field in which QED processes take place.
%
\subsection{The final time-evolved state}
%
We evolve the initial state $|z\rangle$, defined above, in time using the QED $S$-matrix operator $S$. The state evolves non-trivially, due to quantum effects, into the asymptotic final state
\be
    \ket{\text{out}} := S\ket{z} \;,
\ee
which packages all information needed to compute asymptotic observables. (Throughout this paper we assume that we are working with a renormalised Lagrangian and $S$-matrix at least up to a suitable order in the Furry expansion.) To investigate this final state we expose the displacement operator in $\ket{z}$, writing $\ket{\text{out}} = S \mathcal{D}[z]\ket{0}$
and then multiply by unity in the form $1=\mathcal{D}[z]\mathcal{D}^{\dagger}[z]$. This allows us to use (\ref{A-shift}) to turn $S$ into $S[A^\text{cl}]$, the QED $S$-matrix in the background $A^\text{cl}$ defined in (\ref{Acl-def})~\cite{Frantz,Kibble:1965zza,Gavrilov:1990qa,Ilderton:2017xbj}:
\be
 \ket{\text{out}}   = \mathcal{D}[z]\mathcal{D}^\dagger[z] S \mathcal{D}[z] \ket{0} =\mathcal{D}[z] S[A^\text{cl}] \ket{0} \;.
\ee
In order to express the final state itself in terms of objects familiar from strong field QED, namely scattering amplitudes in the classical background $A^\text{cl}$, we insert a complete set of asymptotic states $\ket{f}$ between the remaining $\mathcal{D}[z]$ and $S[A^{\text{cl}}]$, which yields 
\be
\begin{split}\label{final-state-expansion}
    \ket{\text{out}} &=
   \sum_f \mathcal{D}[z]\ket{f}\bra{f}S[A^{\text{cl}}]\ket{0}
    =: \mathcal{D}[z] \sum_f \ket{f}\mathcal{M}(0\to f;A^{\text{cl}}) \;.
\end{split}
\ee
Here $\mathcal{M}(0\to f;A^{\text{cl}})$ is, as desired, the $S$-matrix element for transitions, from vacuum to $\ket{f}$, in the background $A^{\text{cl}}$. In the Furry expansion this matrix element would be calculated by treating the background-field dependence \textit{exactly} (incorporated into the asymptotic states of the theory and the particle propagators), but using the usual perturbative expansion with respect to quantised photons that appear as external scattering states or internal lines, where the (naive~\cite{Ritus1,Narozhnyi:1980dc}) coupling remains the fine structure constant.

\subsection{Observables}
We compute observables as expectation values of operators $\hat{O}$ in the final state. Using the representation (\ref{final-state-expansion}) allows us to express these observables in terms of background field amplitudes. There are broadly two cases to consider. First, suppose that the observable of interest acts only on the matter degrees of freedom. Then $\hat{O}$ commutes with the remaining displacement operator in (\ref{final-state-expansion}) and we have
\be
    \bra{\text{out}} \hat{O}\ket{\text{out}} \to \sum_{f',f} \bar{\mathcal{M}}(0\to f';A^{\text{cl}})\bra{f'}\hat{O}\ket{f} \mathcal{M}(0\to f;A^{\text{cl}}) \;. 
\ee
An example observable is the number of pairs produced. In this case the asymptotic free states $\ket{f}$ are eigenstates of $\ket{O}$, so the double sum collapses to a single sum, and all contributions come from cuts of higher loop amplitudes.

The second case is that $\hat{O}$ acts also on photon degrees of freedom, so $\hat{O}\equiv\hat{O}(a^\dagger,a)$ (suppressing spin and momentum labels), then the displacement operator generates additional terms, and we have
\be\label{general-op-expectation}
    \bra{\text{out}} \hat{O}(a^\dagger,a)\ket{\text{out}} \to \sum_{f',f} \bar{\mathcal{M}}(0\to f';A^{\text{cl}})\bra{f'}\hat{O}(a^\dagger+{\bar z},a+z)\ket{f} \mathcal{M}(0\to f;A^{\text{cl}}) \;. 
\ee
An example of such an observable is the waveform, i.e.~the profile function of the leading $1/r$ falloff of the radiated electromagnetic field in the asymptotic future. Following \cite{Cristofoli:2021vyo}, this takes the form
\be\label{eq:waveform-def}
    W_{\mu\nu}(u,{\bf n}) = -\frac{1}{2\pi} \int\limits_0^\infty\!\hat{\ud}\omega\, e^{-i\omega u}\, k_{[\mu}\varepsilon^\sigma_{\nu]} \bra{\text{out}}a_\sigma(k)\ket{\text{out}} 
    +\text{c.c.}
    \Big|_{k=\omega(1,{\bf n})} \;,
\ee
where ${\bf n}$ defines the point of observation on the 2-sphere at infinity (i.e.~it contains two angular degrees of freedom) while $u$ is retarded time. It is clearly sufficient to work with the spectral, or frequency domain, waveform
\be\label{eq:spectral-waveform-def}
    f_{\mu\nu}(\omega,{\bf n}) := k_{[\mu}\varepsilon^\sigma_{\nu]} \bra{\text{out}}a_\sigma(k)\ket{\text{out}} \Big|_{k=\omega(1,{\bf n})} \;.
\ee
We end with a comment that we have not made any background field approximation -- all we have done is taken an initial coherent state and let it evolve in time.  Nevertheless, our observables can still be expressed in terms of familiar background field amplitudes, plus additional terms as in e.g.~(\ref{general-op-expectation}). In the following sections we consider observables, and therefore the background field amplitudes, order by order in the Furry expansion.  
%

\section{Leading order in the Furry expansion}\label{sec:leading}
%
In this and the following sections we construct the final state $\ket{\text{out}}$ explicitly, order by order in the Furry expansion, analyse its structure, and use it to investigate  observables. We begin at zeroth order -- here there is no backreaction on the initial coherent state, but there is nontrivial physics in the form of pair production.
The material in this section is largely straightforward, and to some extent already appears in the literature, in different guises~\cite{Wald:1975kc,Fradkin:1991zq,Gelis:2006yv,Gelis:2015kya, SangPyo}, but our approach is different and the calculations will serve to fix notation and illustrate useful lessons.

To zeroth order in the Furry expansion the only $\ket{f}$ which can contribute to the sums in (\ref{final-state-expansion}) and (\ref{general-op-expectation}) are (scalar) electron-positron pair states. (Spontaneous photon creation can occur only at higher order, see below.)
The resummed form of the series is well-known -- the vacuum evolves to a \emph{squeezed} state~\cite{GerryKnight2004,Fradkin:1991zq}. This state may be parameterised as, writing $\mathcal{O}_F$ to denote the order of our Furry-like expansion,
\be\label{eq:zeroth_state}
    \ket{\text{out}} = \mathcal{D}[z]\, e^{i \mathcal{W}_{\ccd}}\exp\bigg[\int_{p,q} b^\dagger(p)d^\dagger(q) \mathcal{A}_{\pair}(p,q) \bigg]\ket{0} +\mathcal{O}_F(e)\; \; \; \;  
\ee
in which the constant $\mathcal{W}_{\ccd}$ and the kernel $\mathcal{A}_{\pair}$ are fully determined by the $S$-matrix (equivalently the Schr\"odinger equation) and we have introduced the electron and positron creation operators $b^{\dagger}(p)$ and $d^{\dagger}(q)$. We would like, though, to express these quantities directly in terms of amplitudes. This is done by taking the overlap with e.g.~the vacuum $\ket{0}$ and a single pair state $\ket{p,q}$, which gives
\begin{align}
\braket{0|\text{out}} &=e^{i\mathcal{W}_{\ccd}} \;, \label{vacpersist1}
\\
\braket{p,q|\text{out}} &=\mathcal{A}_{\pair}(p,q)\, e^{i\mathcal{W}_{\ccd}} \;. \label{pair-overlap}
\end{align}    
Hence $\mathcal{W}_{\ccd}$ determines the vacuum persistence amplitude (\ref{vacpersist1}) -- it is the one-loop effective action of the field $A^\text{cl}$. Without the initial coherent state, the effective action would be the free fermion bubble, contributing an irrelevant phase that could be divided out. Here, though, $\mathcal{W}_{\ccd}$ can develop an imaginary part, which via the optical theorem indicates the possibility of pair creation -- it is not an irrelevant phase and should not be divided out, at least not in its entirety.

Turning to (\ref{pair-overlap}), the amplitude for production of a single pair is the \emph{product} of the vacuum persistence amplitude with $\mathcal{A}_{\pair}$. The former will appear in all amplitudes we calculate; if we write a given amplitude as a product $e^{i\mathcal{W}}\mathcal{A}$, as in e.g.~(\ref{pair-overlap}), then it is clear that the $\mathcal{A}$s becomes familiar scattering amplitudes in the vacuum limit, as computed using the usual Feynman rules. Hence the $\mathcal{A}$s are the SFQED \emph{diagrams} computed using Furry picture rules, neglecting the vacuum persistence factor:
\begin{align}
    i\mathcal{W}_{\ccd}&\equiv \begin{tikzpicture}[baseline={([yshift=-.65ex]current bounding box.center)}]
\draw[double] (0,0) circle (0.3);
\end{tikzpicture} \\
\mathcal{A}_{\pair}(p,q) &\equiv  \begin{tikzpicture}[baseline={([yshift=-.65ex]current bounding box.center)}]
 \draw[double] (0,.3) -- (.3,.3);
  \draw[double] (0,-.3) -- (.3,-.3);
\draw[double] (0,0.3) arc (90:270:.3);
\end{tikzpicture}
\end{align}
Amplitudes are therefore \emph{not} represented only by the corresponding connected Feynman diagrams -- the vacuum persistance factor is always present.  This is known, see e.g.~\cite{Fradkin:1991zq} (though that source deals mainly with `reduced' amplitudes in which the vacuum persistence factor is divided out!) but is not often encountered in SFQED because of the dominance of the plane wave model, where
there is no spontaneous particle production. As we consider more complicated processes or initial states, more types of disconnected contributions appear, as we will see.

At this order in the Furry picture,  the only non-trivial amplitudes in our theory are the $1\to 1$ amplitudes, the $0\to 2$ pair production amplitude discussed above, and the $2\to0$ pair annihilation amplitude. All other (asymptotic) quantities must be expressible in terms of these -- the vacuum persistence amplitude, for example, is clearly determined (up to a phase) in terms of the $0\to2$ amplitude by unitarity. We consider this example, and others, below.
    
\subsection{Example: Bogoliubov from amplitudes and crossing symmetry}\label{section:Bogoluibov}
For our first example we revisit the well-known Bogoliubov description of scalar pair creation, reformulating it in terms of amplitudes. In the Bogoliubov approach one observes that because the $S$-matrix is essentially a squeezing operator, the time-evolved ladder operators $B(p) = S^\dagger b(p)S$ and $D(p)=S^\dagger d(p)S$ must be related to $b(p)$ and $d(p)$ through a linear transformation of the form
\be\label{Bogol-def}
\begin{split}
    B(p) = \int_{q}\alpha_{pq}
    b(q) +
    \beta_{pq}
         d^\dagger(q) \;, \qquad
    D^\dagger_p = \int_{q} \alpha^\dagger_{pq}d^\dagger(q) +\beta^\dagger_{pq}b(q) \;,
\end{split}
\ee
for kernels $\alpha_{pq}$ and $\beta_{pq}$ -- we use a `matrix' notation for pair momenta arguments, such that $\alpha^\dagger_{pq}\equiv {\bar \alpha}_{qp}$, and so on. The unitarity of the transformation between $b,d$ and $B,D$, implying $b=SB S^\dagger $ etc, gives us the relations 
\be\label{Bog-relations}
    (\alpha^\dagger \alpha)_{pq} - (\beta \beta^\dagger)_{pq} = \delta_{pq} \;,
    \qquad
    (\alpha \beta)_{pq} = (\beta \alpha)_{pq} \;.
\ee
This allow us to easily invert (\ref{Bogol-def}), yielding
\be\label{Bogol-useful}
\begin{split}
    b(p) =
    \int_{q}\alpha^\dagger_{pq} B(q) -
    \beta_{pq}D^\dagger(q) \;, \qquad
    d^\dagger_p = \int_{q} \alpha_{pq}D^\dagger(q) -\beta^\dagger_{pq}B(q) \;,
\end{split}
\ee
which is used in what follows. One must also define, in this approach, a reference state $\ket{\mathbb{0}}=S^\dagger\ket{\text{in}}$. This state is annihilated by $B(p)$ and $D(p)$, but appears to have no relevance to amplitudes because it evolves the initial state \emph{backward} in time. However, by rewriting amplitudes in terms of $B$, $D$ and $\ket{\mathbb{0}}$, one can explicitly relate them to the Bogoliubov coefficients. To illustrate, we can write the $1\to1$ (electron) and pair production amplitudes as
\be
\begin{split}
    \bra{\text{in}}b(q) S b^\dagger(p)\ket{\text{in}} &= \bra{\mathbb{0}}B(q) b^\dagger(p)\ket{\text{in}} \, \\
    \bra{\text{in}}b(p) d(q) S \ket{\text{in}} &= \bra{\mathbb{0}} B(p)D(q) \ket{\text{in}} \;,
\end{split}
\ee
which, using (\ref{Bogol-def}) and (\ref{Bogol-useful}), and writing the $1\to 1$ electron/positron diagrams, momentum $p\to q$, as $\mathcal{A}_{\forwarde}(p,q)$/ $\mathcal{A}_{\forwardp}(p,q)$ respectively, yields the relations,
\be\label{Bog-as-amplitudes}
\begin{split}
    \mathcal{A}_{\forwarde}(p,q) &= (\alpha^\dagger)^{-1}_{qp}\;,
 \qquad
 \Apair{p}{q} = \int_{r} (\alpha^\dagger)^{-1}_{pr}\beta_{rq} \;.
\end{split}
\ee
Note that the vacuum persistence amplitude has dropped out of these expressions, thus $\alpha$ and $\beta$ are expressible in terms of \emph{diagrams}. This is still somewhat abstract; the physics in the Bogoliubov coefficients is more clearly revealed by thinking of them as operators acting on amplitudes, or more specifically \textit{diagrams}. Consider for example writing the $0\to 2$ amplitude as
\be
\begin{split}\label{crossing-1}
    \bra{\text{in}} d(q)b(p)S\ket{\text{in}}
    &=
    \bra{\text{in}} d(q) S B(p)\ket{\text{in}}
    = \bra{\text{in}} d(q)S\int_r \beta_{pr}d^\dagger(r) \ket{\text{in}} \;,
    \\
    &\implies
    \mathcal{A}_{\pair}(p,q) =
    \int_{r} \beta_{pr}\mathcal{A}_{\forwardp}(r,q) \;.
\end{split}
\ee
In this expression, acting with $\beta$ on the incoming positron leg of the $1\to1$ diagram turns it into an outgoing electron -- Bogoliubov $\beta$ enacts \emph{crossing symmetry}.  Similarly, we can show that acting with $\beta$ on the incoming leg of the $1\to1$ electron scattering diagram it into an outgoing positron:
\be
\label{crossing-2}
 \mathcal{A}_{\pair}(p,q) =
    \int_{r} \beta_{rq}\mathcal{A}_{\forwarde}(r,p) \;.
\ee
Further crossing relations for transforming $0\to 2$ into $1\to 1$ amplitudes are similarly found as:
\be
\label{crossing-3}
\begin{split}
    \mathcal{A}_{\forwardp}(p,q) &= \alpha_{pq} - \int_r \beta^\dagger_{pr} \Apair{r}{q} \;, 
\qquad
    \mathcal{A}_{\forwarde}(p,q) ´= \alpha_{qp} - \int_r  \Apair{q}{r}\beta^\dagger_{pr} \;.
\end{split}
\ee
There is an asymmetry in these expressions compared to (\ref{crossing-1})--(\ref{crossing-2}). The additional $\alpha$-dependent term is required because, physically, there is always a $1\to 1$ amplitude\footnote{For subtleties with this statement related to the choice of time coordinate, and lightfront zero modes, see~\cite{Tomaras:2001vs,Ilderton:2023ifn}.}, but there may not be a $0\to 2$ amplitude, depending on whether or not the field can pair-create; if it cannot, then $\beta=0$ and $\alpha^\dagger = \alpha^{-1}$, consistent with (\ref{Bog-relations}) and~(\ref{Bog-as-amplitudes}). Note that these relations are
exact, since the linear transformation between the creation and annihilation operators is determined by the
$S$-matrix. For a discussion of analogous results in a gravitational context, and an application to Hawking radiation, see~\cite{Aoude:2024sve}.

\section{Next to leading order}\label{sec:nlo}

At next-to-leading order (NLO), that is $\mathcal{O}_F(e)$ in the Furry expansion, photon modes can be populated in the final state, contributing to backreaction on the initial coherent state. We need two more diagrams, specifically
\begin{align}
    \begin{tikzpicture}[baseline]
\draw[double] (0,0) circle (0.2);
\draw[draw=blue, smallsnake] (0.2,0) -- (.6,0);
\end{tikzpicture}&\equiv \mathcal{A}_{\tadpl}(k;\sigma) \;, \label{diagram:tadpole}\\
\begin{tikzpicture}[baseline]
\draw[double] (0,.2) arc (90:270:.2);
\draw[double] (0,.2) -- (.2,.2);
\draw[double] (0,-.2) -- (.2,-.2);
\draw[draw=blue, smallsnake] (-.2,0) -- (.2,0);
\end{tikzpicture}&\equiv \mathcal{A}_{\radpr}(p,q;k,\sigma) \;. \label{diagram:triplet} 
\end{align}
The `vacuum emission' diagram (\ref{diagram:tadpole}) describes the generation of photon modes not present in the classical field defined by the initial coherent state~\cite{Gies:2017ygp}. In other words, the average electromagnetic field in the evolved state differs from that expected by solving the vacuum Maxwell equations, due to quantum effects. (For recent work on these ``tadpole diagrams'' in the presence of constant electromagnetic fields see~\cite{GK, Karb, TadScal, TadSpin, Red}.). The radiation diagram (\ref{diagram:triplet}) describes the production of a pair along with a photon -- one of the physical effects encoded here is simply the radiation of a created pair when it is accelerated by the field of the coherent state.

Though working only to NLO, it is convenient to write contributions to the time-evolved state up to this order in exponential form as
\be
\label{O1-state}
\begin{split}
    \ket{\text{out}} &=
    \mathcal{D}[z] \,
    e^{i\mathcal{W}_{\ccd}}
   \exp\bigg[
        \int_k  \mathcal{A}_{\tadpl}(k;\sigma)a_{\sigma}^{\dagger}(k)
        \bigg] \\
    & \exp\bigg(
        \int_{p,q} b^{\dagger}(p)d^{\dagger}(q)
        \bigg[\mathcal{A}_{\pair}(p,q) +\int_k
        \mathcal{A}_{\radpr}(p,q;k,\sigma)a_\sigma^\dagger(k)\bigg]
    \bigg)
    \ket{0}    +\mathcal{O}_F(e^2) \;,
\end{split}
\ee
which is formally correct to $\mathcal{O}_F(e)$. (See e.g.~\cite{Damgaard:2021ipf,Damgaard:2023ttc} for investigations of `the exponential S-matrix'.) The structure of the state so written is as follows. First, there is no correction to the vacuum persistence amplitude to this order, so $\braket{0|\text{out}}$ is still given by (\ref{vacpersist1}), which appears explicitly in the first line of (\ref{O1-state}) along with an tadpole contribution which describes the possibility of vacuum emission. This resembles a coherent state operator, which we return to in a moment. In the second line of (\ref{O1-state}), we see a squeezing operator as at LO, describing pair production, except that the associated squeezed state profile receives an \emph{operator-valued} shift: this elegeantly captures the fact that produced pairs can now radiate photons.

To see this even more clearly, we point out that the $\mathcal{O}_{F}(e)$ state can be arrived at from the $\mathcal{O}_{F}(e^0)$ state by making the formal shift $\bar{z}_{\sigma}(k)\rightarrow \bar{z}_{\sigma}(k)+a^{\dagger}_{\sigma}(k)$ and keeping only \emph{linear} corrections in $a_\sigma^\dagger$ in the exponent, which is equivalent to retaining one power of $e$. This shift induces the changes
\begin{align}
i\mathcal{W}_{\ccd}\rightarrow i\mathcal{W}_{\ccd}+\int_{k}i\frac{\delta \mathcal{W}_{\ccd}}{\delta \bar{z}^{*}_{\sigma}(k)}a^{\dagger}_{\sigma}(k)
=i\mathcal{W}_{\ccd}+\mathcal{A}_{\tadpl}(k,\sigma)a^{\dagger}_{\sigma}(k)
\,,\\
\mathcal{A}_{\pair}(p,q)\rightarrow \mathcal{A}_{\pair}(p,q)+\int_{k}\frac{\delta \mathcal{A}_{\pair}(p,q)}{\delta \bar{z}^{*}_{\sigma}(k)}a^{\dagger}_{\sigma}(k)
=\mathcal{A}_{\pair}(p,q)+\mathcal{A}_{\radpr}(p,q;k,\sigma)a^{\dagger}_{\sigma}(k)
\,,
\end{align}
in which differentiation of e.g.~the vacuum persistence loop with respect to $z$ essentially pulls out one photon, turning it into the tadpole. This results in the \emph{operator-valued} shifts in the state.

To be concrete, consider the amplitude for the production of a pair, momenta $p$, $q$, and a photon, momentum $\ell$, helicity $s$; assuming the photon is not scattered into the same momentum space as the beam, i.e.~$z_s(\ell)=0$, we have
\be\label{photon-out}
    \bra{\text{in}}b(p)d(q)a_s(\ell)S\ket{\text{in}} = e^{i\mathcal{W}^{(0)}_{\ccd}} \Big( \mathcal{A}_{\radpr}
    {(p,q;\ell,s)}  + \mathcal{A}_{\tadpl}{(\ell;s)} \Apair{p}{q}\Big) \;.
\ee
The radiation diagram appears in the first term. The second term is a \textit{disconnected} contribution in which the pair is produced without radiation, while the photon comes from the vacuum emission process already discussed above. Note that both terms in (\ref{photon-out}) are of the same order in the Furry expansion, but of different loop orders.  Similar statements apply to other processes -- the $\mathcal{O}_F(e)$ diagrams contributing to nonlinear Compton scattering (photon emission from an electron), for example, are the connected three-point diagram, and the disconnected product of the $1\to 1$ diagram with the tadpole diagram:
\be
\begin{tikzpicture}[baseline]
\draw[draw=blue, smallsnake] (0.2,-0.5) -- (1,0.3);
\draw[double] (-0.3,-0.5) -- (1,-0.5);
\end{tikzpicture}
\quad
\raisebox{-5pt}{+}
\quad 
\begin{tikzpicture}[baseline]
\draw[double] (0,0) circle (0.3);
\draw[draw=blue, smallsnake] (0.3,0) -- (1,0);
\draw[double] (-0.3,-0.5) -- (1,-0.5); 
\end{tikzpicture}\;\quad.
\ee
Returning to the structure of the state (\ref{O1-state}), a simple reorganisation of terms takes the emphasis away from squeezing and onto changes to the initial coherent state, i.e.~backreaction on the state.
To this end, we transport the displacement operator $\mathcal{D}[z]$ back toward $\ket{0}$, writing the final state as
\be
\label{O1-state-shifted-D}
\begin{split}
    \ket{\text{out}} &=
    e^{i\mathcal{W}^{(0)}_{\ccd}}
   \exp\bigg[
            \int_{p,q} b^{\dagger}(p)d^{\dagger}(q)\mathcal{A}_{\pair}(p,q) 
        \bigg]
   \exp \bigg[
            \int_k \mathcal{B}(k;\sigma)\big(a_{\sigma}^{\dagger}(k)-{\bar z}_\sigma(k)\big)
        \bigg]\ket{z}
\end{split}
\ee
where 
\begin{align}\label{33}
    \mathcal{B}(k;\sigma)=\mathcal{A}_{\tadpl}(k;\sigma)+\int_{p,q} \mathcal{A}_{\radpr}(p,q;k,\sigma)b^{\dagger}(p)d^{\dagger}(q) \;.
\end{align}
The state has the form $\hat{O}\ket{z}$, in which $\hat{O}$ is an operator in the asymptotic future, which we can interpret as parameterising the deformation of the initial state due to quantum effects. In particular, the final exponential factor in (\ref{O1-state-shifted-D}) takes the form of a displacement operator, with operator-valued profile $\mathcal{B}$, but \emph{defined with respect to vacuum $\ket{z}$} and acting on $\ket{z}$, hence the appearance of the shifted operator $a^\dagger-{\bar z}$. (Note: the conjugate term $B^{\dagger}(k; \sigma)\big(a_\sigma(k)-z_\sigma(k)\big)$ vanishes acting on $\ket{z}$, while  associated normalisation factor can be dropped as it is higher order in $e$.) In this sense, that same exponential creates (shifted) photon excitations on the background $z$, which sounds like what is expected in background field perturbation theory. However, as we will see in the following section, the shifts can be responsible for interference effects which are missed in the background field approximation.

We can also interpret this from the perspective of the vacuum by noting that the product of operators $\hat{O}D[z]$ can be decomposed into  an operator-valued normalisation factor multiplied by a new displacement operator $D[z']$ where $z'_\sigma = z_{\sigma}(k) + \mathcal{B}(k; \sigma)$, again to first order. In this way the photon sector remains, to this order, a coherent state in the asymptotic future, but becomes augmented by excitations of the matter field. 
Further below we will analyse the functional form of the final state via the waveform.

We comment that the use of an exponential representation of the final state is clearly convenient at this order in the Furry expansion. Questions of how to extend this approach to higher orders could be informed by investigations of exponentiation, and quantum remainders, in eikonal scattering~\cite{PhysRevD.57.3763, Laenen2009,KoemansCollado:2018hss,Heissenberg:2021tzo,Cristofoli:2021jas,Haddad2022,Adamo:2022ooq, Luna:2023uwd,Yuchen}, for a review see~\cite{DiVecchia:2023frv}.

\subsection{Example 1: interference effects in observables}\label{subsec:interference}
%
Given the final state (\ref{O1-state}), it is straightforward to obtain information about observables to $\mathcal{O}_F(e)$.  We begin with the spectral waveform (\ref{eq:spectral-waveform-def}), the field-dependent piece of which is contained in the expectation value of the photon mode operator. Using (\ref{O1-state}) we immediately obtain
\be\label{eq:waveform-general-calc-1}
    \bra{\text{out}}a_\sigma(k)\ket{\text{out}} 
    =
    z_\sigma(k)
    +
    \mathcal{A}_{\tadpl}(k;\sigma)
    +
    \int_{p,q} {\mathcal{A}}_{\radpr}(p,q,k;\sigma)
        \bra{\text{out}}b^\dagger(p)d^\dagger(q)\ket{\text{out}} \;.
\ee
The first term is the $O(e^0)$ result that the waveform is just that of the original coherent state -- the initial field evolves according to the classical Maxwell's equations in vacuum. The second and third terms correct this through, respectively, vacuum emission and the radiation produced by created pairs. The structure of the final term is particularly interesting; the \emph{amplitude} for radiation from a pair with momenta $\{p,q\}$ appears integrated against the \emph{expectation value} of the pair creation operator with respect to the final state; this is because any number of pairs can be produced at order $\mathcal{O}_F(e^0)$, which may then go on to radiate. Evaluating the expectation value is simplified, along with calculations and presentation below, by adopting and expanding on the `matrix' notation described after (\ref{Bogol-def}), applied to all \emph{pair} momenta indices. To illustrate:
\be\label{matrix-sewing-def}
    \int_r \Apair{p}{r} 
    \bar{\mathcal{A}}_{\pair} (q,r)
    = \int_r \Apair{p}{r} \Apair{r}{q}^\dagger = (\mathbfbs{\mathcal{A}}_{\pair}\mathbfbs{\mathcal{A}}^{\dagger}_{\pair})_{pq} \;.
\ee
With this, we can complete the evaluation of (\ref{eq:waveform-general-calc-1}) to write the waveform entirely in terms of amplitudes:
\be
\begin{split}\label{big-Z-def}
    \bra{\text{out}}a_\sigma(k)\ket{\text{out}}
    &=
    z_\sigma(k)
    +
    \mathcal{A}_{\tadpl}(k;\sigma)
    +
    \textrm{Tr}\Big[
    \mathbfbs{\mathcal{A}}_{\radpr}(k;\sigma)\mathbfbs{\mathcal{A}}^{\dagger}_{\pair}(1-\mathbfbs{\mathcal{A}}_{\pair}\mathbfbs{\mathcal{A}}^{\dagger}_{\pair})^{-1} \Big]\\
    &=: z_\sigma(k) + \delta z_\sigma(k) \;.
\end{split}
\ee
The structure of the radiation term is the most complicated. It receives contributions from diagrams comprised of an arbitrary numbers of \emph{disconnected} pair production diagrams, but where (at this order) only one of which produces radiation, while the rest contribute a non-trivial factor. This highlights the importance of disconnected contributions in strong field regimes.

A purely background field calculation would not recover the leading term in (\ref{eq:waveform-general-calc-1}), which is not surprising. It has, however, non-trivial consequences at higher orders and for other observables. We can illustrate this using the average number of the produced photons $N_\gamma$, easily found to be
\begin{align}\label{N-gamma-O-e}
    \braket{N_{\gamma}}
    =
    \int_k\braket{\text{out}|\, a^{\dagger}_{\sigma}(k)a_{\sigma}(k)|\text{out}} = \int_k  {\bar z}_\sigma(k)z_\sigma(k) + \Big[{\bar z}_\sigma(k)\delta z_\sigma(k) + z_\sigma(k)\delta {\bar z}_\sigma(k)\Big] + \mathcal{O}_F(e^2) \;.
\end{align}
Beyond the leading order term, we find the same $\delta z_\sigma(k)$ as appears in the waveform, which appears through the interference of $O(e^0)$ and $O(e^1)$ amplitudes. As such, the information contained in $\delta z$ is no longer `painted on the sky' as in the waveform, but is integrated against the initial coherent state profile, only contributing when photons are emitted, via vacuum emission or radiation from pairs, into the same part of momentum space as occupied by the initial coherent state. Such terms are missed in the background field approach, as is easily demonstrated by comparison with $\langle N_\gamma \rangle$ calculated therein, which takes the schematic form
    \be
     \langle N_\gamma \rangle \overset{\text{background}}{\sim} \sum_n n |\mathcal{A}(0\to n)|^2 \;,
    \ee
i.e.~it comes entirely from background field amplitudes mod-squared. The leading contribution is thus at $\mathcal{O}_F(e^2)$ (coming from the mod-squared tadpole and radiation amplitudes ~\cite{Fradkin:1991zq}). While it is again unsurprising that the background-field approach would not count the $\mathcal{O}_F(e^0)$ number of photons in the initial coherent state, we see clearly the consequence that the non-trivial $\mathcal{O}_F(e)$ interference terms in (\ref{N-gamma-O-e}) are also missed.

We can view these interference terms as generalised colinear contributions. Practically, their observation would be difficult in any strong field experiment: for example, most detector equipment placed in the path of a high intensity laser would be severely ablated, making precision measurements challenging. The waveform and other `off-axis' observables are therefore more interesting, from an experimental point of view. 

\subsection{Example 2: all-orders waveform in the impulsive limit}
%
As a concrete example of the formalism outlined above, we study here the example
\begin{align}\label{impulsive-gauge}
    eA_\mu = \delta_\mu^0\delta(x^0)\,  \delta p\cdot x \;, 
\end{align}
in which $\delta p_{\mu}$ is a space-like momentum vector. This potential generates a time-dependent electric field with delta function temporal profile and polarisation in the direction of $\delta p_\mu$. It can be obtained from a finite-duration pulse, e.g.~a Sauter pulse, by taking the short-duration, high-field limit, see~\cite{Ilderton:2019vot}. We will for the field (\ref{impulsive-gauge}) be able to explicitly calculate the pair production and photon emission diagrams and amplitudes above.

As usual, all information on the system's dynamics at leading order in the Furry expansion is contained in the positive/negative energy scalar mode functions of the Klein-Gordon equation in the background.
Thanks to the simple form of the background, the system is integrable at the leading order, we can construct explicitly two sets of mode functions corresponding to free modes in either the asymptotic past ($in$-modes) or asymptotic future ($out$-modes). The $in$-modes are given by 
\begin{align}\label{in-modes_+}
    \phi^{+}_{(in)\mathbf{p}}(x)
    &=
    \Theta(-x^0){e^{-ip\cdot x}} +e^{-i P_{i}x^{i}}\Theta(x^0)
    \sqrt{\frac{p_0}{P_0}}
    \Big(
    \tilde{\alpha}_{\mathbf{p}}{e^{-iP_0 x^0}}
    -\tilde{\beta}_{\mathbf{p}}{e^{+iP_0 x^0}}
    \Big) \;,\\
    \label{in-modes_-}
\phi^{-}_{(in)-\mathbf{p}}(x)
    &=
    \Theta(-x^0) {e^{ip_0 x^0-ip_ix^i}}
    +
    e^{-i P_{i}x^{i}}\Theta(x^0)
    \sqrt{\frac{p_0}{P_0}}\Big(
    \tilde{\alpha}_{\mathbf{p}}{e^{iP_0 x^0}}-\tilde{\beta}_{\mathbf{p}}{e^{-iP_0 x^0}}
    \Big) \;,
\end{align}
where $p_0=\sqrt{\mathbf{p}^2+m^2}$, $(P_{0},P_i)=(\sqrt{(\mathbf{p}+\delta\mathbf{p})^2+m^2},p_i+\delta p_i)$ and
\begin{align}
    \tilde{\alpha}_{\mathbf{p}}=\frac{1}{2}\left(\sqrt{\frac{p_0}{P_0}}+\sqrt{\frac{P_0}{p_0}}\right)\;,
    \qquad
    \tilde{\beta}_{\mathbf{p}}=\frac{1}{2}\left(\sqrt{\frac{p_0}{P_0}}-\sqrt{\frac{P_0}{p_0}}\right)
\end{align}
The $out$-modes are given by
\begin{align}\label{eq:phi_out_plus}
    \phi^{+}_{(out)\mathbf{P}}(x)
    &=
    \Theta(x^0){e^{-iP\cdot x}} +
    e^{-i p_{i}x^{i}}\Theta(-x^0)
    \sqrt{\frac{P_0}{p_0}}\Big(
    \tilde{\alpha}_{\mathbf{p}}{e^{-ip_0 x^0}}+\tilde{\beta}_{\mathbf{p}}{e^{ip_0 x^0}}
    \Big) \;,\\
    \label{eq:phi_out_minus}
    \phi^{-}_{(out)-\mathbf{P}}(x)
    &=
    \Theta(x^0){e^{iP_0 x^0-iP_ix^i}} +e^{-i p_{i}x^{i}}\Theta(-x^0)
    \sqrt{\frac{P_0}{p_0}}\Big(
    \tilde{\alpha}_{\mathbf{p}}{e^{ip_0 x^0}}+\tilde{\beta}_{\mathbf{p}}{e^{-ip_0 x^0}}
    \Big) \;,
\end{align}
where, here and below, $\mathbf{P}:=\mathbf{p}+\delta \mathbf{p}$. As usual, the scalar field can be expanded using either set of modes,
\begin{align}\label{eq:Phi_in}
    \Phi(x)&=
    \int_p\,
    \left[b(p)\phi^{+}_{(in)\mathbf{p}}(x)+d^{\dagger}(p)\phi^{-}_{(in)\mathbf{p}}(x)\right]\\
    \label{eq:Phi_out}
    &=
    \int_p\,
    \left[B(p)\phi^{+}_{(out)\mathbf{p}}(x)+D^{\dagger}(p)\phi^{-}_{(out)\mathbf{p}}(x)\right] \;,
\end{align}
where the operators $B,D$ are those introduced in Sec.~\ref{section:Bogoluibov}. To see this, we use the Heisenberg evolution of the field operator to write
\begin{align}\label{eq:Phi_Heisenberg}
    \lim_{t\to\infty}\Phi(t,\mathbf{x})=\lim_{t\to-\infty} S^{\dagger}\Phi(t,\mathbf{x})S \;,
\end{align}
and then compare with \eqref{eq:Phi_in} and \eqref{eq:Phi_out}, where the definition of the in and out modes yields the asymptotic behaviour
\begin{align}
    \lim_{t\to-\infty}\Phi(t,\mathbf{x}) \sim \int_p\,
    b(p)e^{-ip\cdot x}+d^{\dagger}(p)e^{ip\cdot x}\;,
    \qquad
    \lim_{t\to \infty}\Phi(t,\mathbf{x})\sim\int_p\,
    B(p)e^{-ip\cdot x}+D^{\dagger}(p)e^{ip\cdot x}\,.
\end{align}
It follows directly that $B(p)=S^{\dagger}b(p)S$ and $D^{\dagger}(p)=S^{\dagger}d^{\dagger}(p)S$, as previously.

The final ingredient needed to compute amplitudes is the scalar propagator. The $in$-$out$ Feynman propagator $G_{+-}(x',x)$ (i.e.~that which naturally arises in an amplitudes context) is
\begin{align}
    G_{+-}(x',x)=\frac{\braket{\mathbb{0}|T\bar{\Phi}(x')\Phi(x)|\textrm{in}}}{\braket{\mathbb{0}|\textrm{in}}}    
\end{align}
Expanding the fields in terms of modes as above, we have
\begin{align}
    G_{+-}(x',x)=
    \int_p
    \sqrt{\frac{p_0}{P_0}}\frac{1}{\tilde{\alpha}_{\mathbf{p}}}
    \left[\Theta(x'^{0}-x^0)\bar{\phi}^{-}_{(out)-\mathbf{P}}(x')\phi^{-}_{(in)-\mathbf{p}}(x)+\Theta(x^{0}-x'^0)\bar{\phi}^{+}_{(in)\mathbf{p}}(x')\phi^{+}_{(out)\mathbf{P}}(x)\right]
\end{align}
We will find in our calculations below that the $in$-$in$ propagator $G_{--}(x',x)$ also arises. It is similarly defined by
\begin{align}
G_{--}(x',x)&=\braket{\textrm{in}|T\bar{\Phi}(x')\Phi(x)|\textrm{in}}\\
&=\int_p\left[\Theta(x'^{0}-x^0)\bar{\phi}^{-}_{(in)-\mathbf{p}}(x')\phi^{-}_{(in)-\mathbf{p}}(x)+\Theta(x^{0}-x'^0)\bar{\phi}^{+}_{(in)\mathbf{p}}(x')\phi^{+}_{(in)\mathbf{p}}(x)\right]
\end{align}

\subsubsection{Diagrams for $1 \to 1$ and $0\to 2$ processes}
The Bogoluibov coefficents $\alpha_{pq}$ and $\beta_{pq}$ can be computed explicitly for our example. To do so, we rewrite \eqref{in-modes_-} and \eqref{in-modes_+} as, adopting the compact notation $\hat{\ud}p=\ud p/2\pi$ and $\hat{\delta}(p)=2\pi\delta(p)$,
\begin{align}
   \phi^{+}_{(in)\mathbfbs{p}}(x)&=\int_{q}\tilde{\alpha}_{\mathbf{p}}(2q_0){\hat{\delta}^3}
   (\mathbf{p}+\delta \mathbf{p}-\mathbf{q})\phi^{+}_{(out)\mathbfbs{q}}(x)-\int_{q}\tilde{\beta}_{\mathbf{p}}(2q_0){\hat{\delta}^3}(\mathbf{p}+\delta \mathbf{p}+\mathbf{q})\phi^{-}_{(out)\mathbfbs{q}}(x)\\
    \phi^{-}_{(in)\mathbfbs{p}}(x)&=\int_{q}\tilde{\alpha}_{\mathbf{p}}(2q_0){\hat{\delta}^3}(\mathbf{p}-\delta \mathbf{p}-\mathbf{q})\phi^{-}_{(out)\mathbfbs{q}}(x)-\int_{q}\tilde{\beta}_{\mathbf{p}}(2q_0){\hat{\delta}^3}(\mathbf{p}+\delta \mathbf{p}+\mathbf{q})\phi^{+}_{(out)\mathbfbs{q}}(x) \;.
\end{align}
Substituting into \eqref{eq:Phi_in} and comparing with \eqref{eq:Phi_out}, we find 
\begin{align}
    \alpha_{pq}=\tilde{\alpha}_{\mathbf{p}}(2q_0){\hat{\delta}^3}(\mathbf{p}+\delta \mathbf{p}-\mathbf{q});\qquad
 \beta_{pq}=-\tilde{\beta}_{\mathbf{p}}(2q_0){\hat{\delta}^3}(\mathbf{p}+\delta \mathbf{p}+\mathbf{q}).
\end{align}
It follows, using the general expressions (\ref{Bog-as-amplitudes}) that the diagrams for $1 \rightarrow 1$ electron/positron scattering $\mathcal{A}_{\forwarde}(p,q)$/ $\mathcal{A}_{\forwardp}(p,q)$ and pair creation $\mathcal{A}_{\pair}$ at leading order in the Furry expansion of observables are
\begin{align}
    \mathcal{A}_{\forwarde}(p,q)
    &=
    \frac{1}{\tilde{\alpha}_{\mathbf{q}}}
    (2q_0){\hat{\delta}^3}(\mathbf{p}-\delta\mathbf{p}-\mathbf{q}) \;,
    \qquad 
    \mathcal{A}_{\forwardp}(p,q)
    =
    \frac{1}{\tilde{\alpha}_{-\mathbf{q}}}
    (2q_0){\hat{\delta}^3}(\mathbf{p}+\delta\mathbf{p}-\mathbf{q}) \;,
    \\
\mathcal{A}_{\pair}
&=
-\frac{\tilde{\beta}_{\mathbf{p}}}{\tilde{\alpha}_{\mathbf{p}}}
(2p_0){\hat{\delta}^3}(\mathbf{p}+\mathbf{q}) \;.
\end{align}
From these explicit expressions it is straightforward to verify the crossing symmetry relations~\eqref{crossing-1}--\eqref{crossing-3}.

\subsubsection{Waveform in the time-like asymptotic future}
%
The two irreducible diagrams required for calculating leading backreaction effects are $\mathcal{A}_{\tadpl}$ (the tadpole) and $\mathcal{A}_{\radpr}$ (pair plus radiation). These enter the waveform at future null infinity, (\ref{eq:spectral-waveform-def}), through (\ref{eq:waveform-general-calc-1}). However, the electromagnetic fields of our chosen background (\ref{impulsive-gauge}) are homogeneous in space, which means the current density of created pairs does not have compact support, which violates the usual assumption of asymptotic falloff used in the derivation of (\ref{eq:waveform-def}) and (\ref{eq:spectral-waveform-def}). In our case, we will see instead that the contributing amplitudes are supported only at zero frequency, such that the waveform  (\ref{eq:spectral-waveform-def}) at future \emph{null} infinity actually vanishes. Rather, in our example, information about backreaction is encoded in the electromagnetic field at \textit{time-like} future infinity, so this is the object we will study.

To this end, we note that the retarded field strength $F_{\mu\nu}$ \emph{sourced} by a current $j_\mu$ has the form
\begin{align}\label{eqn:F}
F_{\mu\nu}&=
\int\! \hat{\ud}^4k\left(\frac{-2i}{(k_0+i0^{+})^2-\mathbfbs{k}^2}\right)k_{\left[\mu\right.}j_{\left.\nu\right]}(k)e^{-ik\cdot x}\,.    
\end{align}
Identifying $F_{\mu\nu}$ with the expectation value $\braket{F_{\mu\nu}}=\braket{\textrm{out}|F_{\mu\nu}|\textrm{out}}$, the relevant current for studying leading-order backreaction is $\delta z_{\mu}(k)$, introduced in (\ref{big-Z-def}), but where $k_\mu$ is taken off-shell. The contributing diagrams $\mathcal{A}_{\tadpl}$ and $\mathcal{A}_{\radpr}$ should then be understood as off-shell currents. Let us compute these for our chosen background.

The tadpole diagram can be computed from the matrix element of the gauge invariant current density operator as
\begin{align}\label{tad-off-1}
    \mathcal{A}_{\tadpl}(k)_{\mu}&=\frac{ie\int \ud^4x\, e^{ik\cdot x} \braket{ \mathbb{0}| T\bar{\Phi}(x)\stackrel{\leftrightarrow}{D}_{\mu}\Phi(x)|\textrm{in} }}{\braket{\mathbb{0}|\textrm{in}}}\\
    &=ie \int \!\ud^4x\, e^{ik\cdot x}\left.\left(D^{(x)}_{\mu}-{\bar D}^{(x')}_{\mu}\right)G_{+-}(x',x)\right\rvert_{x'\rightarrow x}\,,
    \label{tad-off-1-5}
\end{align}
where, throughout this paper, the coincidence limit of operator products is taken using a symmetric point-splitting prescription. Since the particle wavefunctions and propagator involve only elementary functions, it is straightforward to perform the Fourier transform in (\ref{tad-off-1-5}). One finds the non-zero components
\begin{align}\label{tad-off-2}
    \mathcal{A}_{\tadpl}(k)_{i}&=\hat{\delta^3}(\mathbf{k})(ie)\int_p(-2 p_i)\left(\frac{1}{k_0-2p_0+i0^{+}}+\frac{1}{k_0+2p_0-i0^{+}}\right)\frac{\tilde{\beta}_{\mathbf{p}}}{\tilde{\alpha}_{\mathbfbs{p}}} \;,
\end{align}
where the spatial delta function arises due to the homogeneity of the background.

Moving on, it is convenient to compute $\mathcal{A}_{\radpr}$ by writing it as the difference between the entire amplitude for photon emission, and the \textit{disconnected} contribution to the same, thus: 
\begin{align}\label{del1}
   \mathcal{A}_{\radpr}(p,q,k)_{\mu}&=\frac{\braket{\mathbb{0}|B(p)D(q)\left[ie \int\!\ud^4x\, e^{ik\cdot x}
   \,T\bar{\Phi}(x)\stackrel{\leftrightarrow}{D}_{\mu}\Phi(x)
   \right]|\textrm{in} }}{\braket{\mathbb{0}|\textrm{in}}}-\mathcal{A}_{\pair}(p,q)\mathcal{A}_{\tadpl}(k)_{\mu}\,.
\end{align}
This can be simplified, using expressions~\eqref{eq:Phi_in} and~\eqref{eq:Phi_out} for the scalar field, to
\begin{align}
   \mathcal{A}_{\radpr}(p,q,k)_{\mu}&= \frac{ie}{\tilde{\alpha}_{\mathbf{p}}\tilde{\alpha}_{-\mathbf{q}}}
   \int\!\ud^4x\, e^{ik\cdot x}\bar{\phi}^{+}_{(in)\mathbf{p}}(x)\stackrel{\leftrightarrow}{D}_{\mu}\phi^{-}_{(in)\mathbf{q}}(x) \;.
\end{align}
Inserting this into the quantity of interest for computing $\delta z_\mu$, we have
\begin{equation}\label{del2}
    \textrm{Tr}\left[(1-\mathbfbs{\mathcal{A}_{\pair}}\mathbfbs{\mathcal{A}^{\dagger}_{\pair}})^{-1}\cdot \mathbfbs{\mathcal{A}_{\radpr}}\mathbfbs{\mathcal{A}^{\dagger}_{\pair}}\right]_{\mu}
    =-ie\int \!\hat{\ud}\mathbf{p}\frac{\tilde{\beta}_{\mathbf{p}}}{\tilde{\alpha}_{\mathbf{p}}}\int d^4x\,e^{ik\cdot x}\bar{\phi}^{+}_{(in)\mathbf{p}}(x)\stackrel{\leftrightarrow}{D}_{\mu}\phi^{-}_{(in)-\mathbf{p}}(x)\,.
\end{equation}
Combining the two pieces (\ref{del1}) and (\ref{del2}), we can determine the off-shell analog of $\delta z_{\sigma}$ in (\ref{big-Z-def}) as
\begin{equation}
    \delta z_{\mu}=ie \int \!\ud^4x\, e^{ik\cdot x}\left.\left(D^{(x)}_{\mu}-{\bar D}^{(x')}_{\mu}\right)G_{--}(x',x)\right\rvert_{x'\rightarrow x}
    \eqqcolon \hat{\delta^3}(\mathbf{k})j_{\mu}(k_0)\,,
\end{equation}
which, we note, is precisely the current one finds in the `in-in' formalism, i.e.~the mean current generated by pair production. It is interesting that to obtain this from `in-out' quantities, i.e.~amplitudes, one has to combine (\ref{del1}) and (\ref{del2}) which, formally, contains an infinite sum over in-out diagrams, through the inverse operator appearing in the trace. This is a point we return to in equation \eqref{eq:waveform_resumm} of Sec.~\ref{subsec:photon_no_NLO}.

We find, for the non-vanishing parts of the current
\begin{align}\label{eqn:j_0}
    j_{i}(k_0)= -ie \int_p \tilde{\alpha}_{\mathbfbs{p}}\tilde{\beta}_{\mathbfbs{p}}(2p_i)\left(\frac{1}{k_0+2p_0+i0^{+}}+\frac{1}{k_0-2p_0+i0^{+}}\right) \;,
\end{align}
For a current of the form (\ref{eqn:j_0}), $F_{\mu\nu}$ reduces to one of a temporally dependent electric field.
Inserting~(\ref{eqn:j_0}) into~(\ref{eqn:F}), and performing the $k_0$ integral we pick-up contributions from the poles at $k_0=0$ and, via $j_\mu$, poles at $k_0 = \pm 2p_0$. The non-zero components of the resulting electromagnetic field are
\begin{align}
    F_{0i} &=\textrm{Im} \int \frac{\ud^3\mathbfbs{p}}{(2\pi)^3(\mathbf{p}^2+m^2)} (e\tilde{\alpha}_{\mathbfbs{p}}\tilde{\beta}_{\mathbfbs{p}})p_ie^{2i\sqrt{\mathbf{p}^2+m^2}x^0} \;.
\end{align}
The angular integral  can be performed exactly, yielding 
\be
    F_{0i} = \textrm{Im}\,
   \frac{\delta p_i}{|\delta p|}\int_{0}^{\infty}\!\ud p\, f(p)\, e^{2ix^0\sqrt{p^2+m^2}} \;,
\ee
in which $f(p)$ has a lengthy and unilluminating form. While the remaining integral over $p\equiv|{\bf p}|$ is involved, we fortunately only need to evaluate it in a saddle point approximation in order to read off the leading asymptotic behavior of $F_{\mu\nu}$ as $x^0\rightarrow \infty$. For this we only need the $p\rightarrow 0$ behavior of $f(p)$, which is explicitly given by
\begin{align}
    f(p)=
    -\frac{e|\delta {\bf p}| m  \left(\delta p^2+2 m^2\right)}{{24} \pi ^2  \left(\delta p^2+m^2\right)^{3/2}}
    \bigg(\frac{p}{m}\bigg)^4+\mathcal{O}(p^5/m^5)\,.    
\end{align}
Finally, the leading asymptotic behaviour of the \emph{back-reacted} electromagnetic field 
can be computed via the saddle-point approximation,
and is given  by the fairly simple expression
\begin{align}
    F_{\mu\nu}=
    (\ud x^0\wedge \delta p)_{\mu\nu}\left[\frac{e m^2 \left(\delta p^2+2 m^2\right) }{{4} (2\pi) ^{3/2} \left(\delta p^2+m^2\right)^{3/2}}\right]\frac{\sin \left(2 m x^0+\frac{\pi }{4}\right)}{(2m x^0)^{5/2}}+\mathcal{O}((m x^0)^{-7/2})\,.
\end{align}
We see that the asymptotic back-reacted field is a damped oscillator satisfying $F''(x_0)+\gamma(x^0)F'(x_0)+\omega^2(x^0)F(x_0)=0$, where the time-dependent frequency and damping are respectively $\omega^2(x^0)=\tfrac{15}{4(x^0)^2}+4m^2$ and $\gamma(u)=\tfrac{5}{x^0}$.

\section{Resummation}\label{sec:resum}
In this section we explore several observables as, at each order in the Furry expansion, resummations of pair creation diagrams. This will expose a great deal of structure. 

\subsection{Vacuum persistence}\label{subsec:vp1}
We begin by revisiting the vacuum persistence probability $\mathbb{P}$, i.e.~the probability of creating no pairs, at leading order in the Furry expansion. This is, from \eqref{O1-state},
\begin{align}
    \mathbb{P}=\left|\braket{\textrm{in}|\textrm{out}}\right|^2
    =\left|e^{i\mathcal{W}_{\ccd}}\right|^2=e^{-2\textrm{Im}\, \mathcal{W}_{\ccd}}\,.
\end{align}
Unitarity, in the form $\braket{\text{out}|\text{out}}=1$, implies that the same probability can be expressed via
\begin{align}\label{eq:no_pair_is_exp_A}
 e^{2\text{Im}\,\mathcal{W}_{\ccd}} = 
 \left|\exp\bigg[\int_{p,q} b^\dagger(p)d^\dagger(q) \mathcal{A}_{\pair}(p,q) \bigg]\ket{\textrm{in}}\right|^{2}\,.
\end{align}
Our task is to evaluate the right-hand side of this expression explicitly. To keep track of the powers of $\mathcal{A}_{\pair}$ which arise when expanding the exponential in \eqref{eq:no_pair_is_exp_A}, let us define the function
\begin{align}\label{b-def}
    \mathbb{b}(t)=\left|\exp\bigg[t\int_{p,q} b^\dagger(p)d^\dagger(q) \mathcal{A}_{\pair}(p,q) \bigg]\ket{\textrm{in}}\right|^2\,,
\end{align}
such that $\mathbb{P}=\mathbb{b}(1)^{-1}$. The first few terms in the Taylor series of $\mathbb{b}(t)$ are easily found as
\begin{align}\label{b}
\mathbb{b}(t)&
    =1
    +
    t\underbrace{\textrm{Tr}\, \boldpair\boldpair^\dagger}_{\begin{tikzpicture}[baseline]
\draw[double] (0,0) circle (0.3) ;
\draw[dashed] (0,0.35)--(0,-0.35);
\end{tikzpicture}}
+
\frac{t^2}{2!}
\Bigg[\,
\underbrace{\left(\textrm{Tr}\, \boldpair\boldpair^\dagger\right)^2}_{\left(\begin{tikzpicture}[baseline]
\draw[double] (0,0) circle (0.5*3/5) ;
\draw[dashed] (0,0.6*3/5)--(0,-0.6*3/5);
\end{tikzpicture}\right)^2
}
+
\underbrace{\mathrm{Tr}\,\boldpair\boldpair^\dagger\boldpair\boldpair^\dagger}_{\begin{tikzpicture}[baseline]
\draw[double] (0,0.5*3/5) arc (90:270:0.25*3/5) ;
\draw[double] (0,0) arc (90:-90:0.25*3/5); 
\draw[double] (0,-.5*3/5) arc (90:270:0.25*3/5) ; 
\draw[double] (0,-1*3/5) arc (-90:90:0.75*3/5); 
\draw[dashed] (0,0.6*3/5)--(0,-1.2*3/5);
\end{tikzpicture}}
\,\Bigg]
+\ldots
\end{align}

We have included the diagrammatic representation of each term -- the dashed vertical lines are cuts, across which particle lines are sewn by on-shell momentum integrals (recall~(\ref{matrix-sewing-def})). There is interesting structure already at order $t^2$ in this expansion: using the diagrammatic representation, it is clear that the two terms present saturate the distinct ways in which one can sew the product of \emph{two} pair-production (horseshoe) diagrams to their conjugates.

Proceeding to higher orders in the expansion of (\ref{b-def}) we will obtain a representation of $\mathbb{P}^{-1}$ as a sum of all cut diagrams contributing to the probability of creation of all numbers of pairs. The coefficient of $t^n$ in this series will be the un-normalised probability for the creation of $n$ pairs. These higher orders terms quickly become rather complicated, though, so to conveniently organise diagrams contributing to the coefficient of $t^n/n!$, we define the cut diagram $\mathcal{B}(p,p')$ by
\begin{align}
   \mathcal{B}(p,p'):=  \int_{q} \mathcal{A}_{\pair}(p,q)\bar{\mathcal{A}}_{\pair}(p',q) =\begin{tikzpicture}[baseline]
\draw[double] (-.2,.25) arc (-250:60:.5) ;
\draw[dashed] (0,-.25)--(0,-1);
\draw (-.2,.45) node {$p$};
\draw (.25,.5) node {$p'$};
\end{tikzpicture}\,,
\end{align}
and we will again adopt a matrix representation, writing $\mathbfbs{\mathcal{B}}_{pp'}=\mathcal{B}(p,p')$. With this notation, we can rewrite and extend the expansion (\ref{b}) of $\mathbb{b}(t)$ as
 \begin{align}
    \mathbb{b}(t)=
    1
    +
    t\,\textrm{Tr}\, \mathbfbs{\mathcal{B}}
    +
    \frac{t^2}{2!}
    \Big(
    (\textrm{Tr}\, \mathbfbs{\mathcal{B}})^2+\textrm{Tr}\, (\mathbfbs{\mathcal{B}}^2)
    \Big)
    +...+\mathbb{b}_{n} \frac{t^n}{n!}+...
\end{align}
It is clear that $\mathbb{b}_{n}$, the coefficient of $t^n/n!$, will be a polynomial in the set of variables $\{\textrm{Tr}(\mathbfbs{\mathcal{B}}^j);1\leq j\leq n \}$, such that the terms take the form
\begin{align}\label{eq:typical_term_in_Pt}
    (\textrm{Tr}\, (\mathbfbs{\mathcal{B}}^{j_1}))^{n_1}
    (\textrm{Tr}\, (\mathbfbs{\mathcal{B}}^{j_2}))^{n_2} \, \ldots\,
   (\textrm{Tr}\, (\mathbfbs{\mathcal{B}}^{j_r}))^{n_r}\quad;\quad \sum_{i}j_{i}n_i=n\,,
\end{align}
where the constraint simply enforces that the number of pairs contributing to $\mathbb{b}_{n}$ is equal to $n$. To completely determine the polynomials $\mathbb{b}_{n}$, however, we also need the exact coefficient of these general terms.

Notice that a typical term of form \eqref{eq:typical_term_in_Pt} translates to a \textit{disconnected} diagram, made of products of the connected pieces $\textrm{Tr}\,\mathbfbs{\mathcal{B}}^{j_i}$. The coefficient of this term, in the polynomial $\mathbb{b}_n$, can be identified with the combinatorial factor counting the number of permutations of the particle/anti-particle legs that give rise to the (topologically) same \textit{disconnected} diagram, which we must identify.

The coefficient of the term shown in \eqref{eq:typical_term_in_Pt} should itself be a product of two factors, (1) the product of the number of ways $n$ can be partitioned into $n_i$ blocks of size $j_i$ and (2) the product of the number of permutations of $\mathbfbs{\mathcal{B}}$ possible in each factor of $\textrm{Tr}(\mathbfbs{\mathcal{B}}^{j_i})$. The former factor is well known to be expressible in terms of the coefficients in the so-called Bell polynomial\footnote{The Bell polynomial can be generated via $B_{n}(x_{1},\ldots ,x_{n})=\left.\left({\frac {\partial }{\partial t}}\right)^{n}\exp \left(\sum _{j=1}^{n}x_{j}{\frac {t^{j}}{j!}}\right)\right|_{t=0}$ see e.g.~\cite{comtet2012advanced}.}. The latter factor is easily found to be $\prod_i(j_i-1)!$. (This is not $\prod_ij_i!$ due to the trace, which is invariant under cyclic permutation.) Combining these two factors, we can write the coefficient of $t^n/n!$ in the form
\begin{align}
    \mathbb{b}_{n}=B_{n}(\textrm{Tr}[\mathbfbs{\mathcal{B}}],\textrm{Tr}[\mathbfbs{\mathcal{B}}^2],2!\textrm{Tr}[\mathbfbs{\mathcal{B}}^3],...,(n-1)!\textrm{Tr}[\mathbfbs{\mathcal{B}}^n])
\end{align}
where, $B_{n}(x_1,...,x_n)$ is the Bell polynomial~\cite{comtet2012advanced},
and where the factors of $(j_i-1)! $ in the arguments of the polynomial account for the permutations discussed above. 

It is interesting to compare to~\cite{Gelis:2006yv}, which considered scalar particle production stemming from a current source. The expressions therein for the sum over $n-$particle cut diagrams
agree with the expressions above, upon expanding the Bell polynomials as $B_n(x_1,\dots,x_n)=\sum_{j_1+\dots +nj_n=n} \prod^n_{i=1}x_i^{j_i}/(i!)^{j_i}j_i!$, and then making the replacement $\textrm{Tr}[\mathbfbs{\mathcal{B}}^j]\to j b_j/g^2$ (we refer readers to~\cite{Gelis:2006yv} for conventions).
That we have the same Bell polynomial structure in LO vacuum persistence, despite considering different models, falls out naturally simply because of the ``combinatorics of grouping [$n$] objects into [$j_i$] clusters"~\cite{Gelis:2015kya}. However, we emphasise that in our approach we are not restricted to LO vacuum persistence -- higher internal loops and backreaction are incorporated into our setup; indeed, we already highlighted this in the context of interference effects in the waveform, and below we will give the NLO corrections to vacuum persistence.
The series expansion for $\mathbb{b}(t)$ can then be written as
\begin{align}\label{eq:b(t)_is_Tr_exp}
    \mathbb{b}(t)=\sum_{n=0}^{\infty}\frac{t^n}{n!}B_{n}\left(\textrm{Tr}[\mathbfbs{\mathcal{B}}],\textrm{Tr}[\mathbfbs{\mathcal{B}}^2],2!\textrm{Tr}[\mathbfbs{\mathcal{B}}^3],...,(n-1)!\textrm{Tr}[\mathbfbs{\mathcal{B}}^n]\right)=\exp\left(\sum_{j=1}^{\infty}t^j\frac{\textrm{Tr}[\mathbfbs{\mathcal{B}}^j]}{j}\right)\,,
\end{align} 
where the last line follows from the definition of $B_{n}(x_1,...,x_n)$~\cite{comtet2012advanced}. The series in the exponent can be further simplified to arrive at
\begin{align}
    \mathbb{b}(t)&=\exp\left(-\textrm{Tr} \log(1-t\mathbfbs{\mathcal{B}})\right) =
    {\det}(1-t\mathbfbs{\mathcal{B}})^{-1} \;.
\end{align}
This reproduces, of course, the expected expression for $\mathbb{P} =\mathbb{b}^{-1}(1)=\det(1-\mathbfbs{\mathcal{B}})\;.$ What is surprising is the depth of structure in this concise expression, when resolved into Feynman diagrams. It also provides a realisation of unitarity by expressing the imaginary part of the effective action in terms of (forward) scattering into pairs  (i.e. optical theorem). Given this, and before moving on to higher orders in the Furry expansion of $\mathbb{P}$, we highlight an interesting corollary of the above.

\subsubsection{Pair distributions}\label{subsub:pair_dist}
The probability of creating exactly $n$ pairs can be written explicitly as
\begin{align}\label{eq:Bell_distribution}
  \mathbb{P}_{n}=  \frac{\det(1-\mathbfbs{\mathcal{B}})}{n!}B_{n}(\textrm{Tr}[\mathbfbs{\mathcal{B}}],\textrm{Tr}[\mathbfbs{\mathcal{B}}^2],2!\textrm{Tr}[\mathbfbs{\mathcal{B}}^3],...,(n-1)!\textrm{Tr}[\mathbfbs{\mathcal{B}}^n])\,,
\end{align}
which we now explore through some special cases. First, as a check, consider replacing $\mathbfbs{\mathcal{B}}$ with a $\mathbb{C}$-number $\mathcal{B}$. Using $B_{n}(0!,1!,2!,...(n-1)!)=n!$ we then correctly recover the statistics of a two-mode squeezed state:
\begin{align}
  \mathbb{P}_{n}\bigg|_{\mathbfbs{\mathcal{B}}\rightarrow \mathcal{B}}= (1-|\mathcal{B}|)|\mathcal{B}|^n\;.
\end{align}
Another special case corresponds to a homogeneous, but time-dependent, field, for which 
\begin{align}
    [\mathbfbs{\mathcal{B}}^j]_{pp'}\rightarrow |f(p)|^{2j} (2p_0)(2\pi)^3\delta^3(\mathbfbs{p}'-\mathbfbs{p})\quad;\quad j=1,2,3...\,,
\end{align}
where $f(p)$ carries the nontrivial information on the pair spectrum. This implies
\begin{align}\label{eq:TrBJ_is_lambdaj}
     \textrm{Tr}(\mathbfbs{\mathcal{B}}^j) \to \lambda_j := V_3\int_p (2p_0)|f(p)|^{2j} \,,
\end{align}
where $V_3=(2\pi)^3\delta^3(0)$ is the spatial volume. In this case, the probability of creating exactly $n$ pairs becomes
\begin{align}\label{pnnBl}
    \mathbb{P}_n&=\frac{1}{n!}\, \exp\left[V_3\int_p(2p_0)\log\left(1-|f(p)|^2\right)\right]\, B_{n} 
    \left(
    \lambda_1, 
    \lambda_2, 
    2! \lambda_3, \ldots \,, 
    (n-1)! \lambda_n 
    \right)\,. 
\end{align}
If the probability of creating a pair is small (as in e.g.~the weak field regime), and we can impose $\lambda_j \ll \lambda_1 \ll 1 \,\forall \, j>1$, then (\ref{pnnBl}) can be approximated by
\begin{align}
    \mathbb{P}_n\approx e^{-\lambda_1}\frac{\lambda_1^n}{n!} \;,
\end{align}
which is the expected Poisson distribution~\cite{Gelis:2006yv,Fukushima:2009er}. To go beyond this approximation and examine corrections to the Poisson distribution we include effects linear in $\lambda_2$. The resulting probability distribution is
\begin{align}\label{Pn-norm}
    \mathbb{P}_n\approx \left(\frac{e^{-\lambda_1}}{1+\frac{\lambda_2}{2}}\right)\frac{\left(\lambda_1^n+\binom{n}{2}\,\lambda_1^{n-2}\lambda_2\right)}{n!} \;,
\end{align}
which can be interpreted as a one-parameter deformation of the Poisson distribution. (We have left the denominator un-expanded so that (\ref{Pn-norm}) is explicitly normalised.)
A non-Poissionian distribution was also found in~\cite{Gelis:2006yv} by studying a variance of the multiplicity distribution and analysing contributing diagrams; a Poissonian distribution could only have resulted if (in our language) $\textrm{Tr}[\mathbfbs{\mathcal{B}}^j]$ for $j>1$ were neglected.
We shall see in Sec.~\ref{sec:NLO_vac_to_vac} that, at NLO, it is \textit{necessary} to consider the deformed distribution \eqref{Pn-norm} to model all the physical effects that arise at $\mathcal{O}_F(e^2)$. This result has potential practical applications in the benchmarking of numerical strong field QED schemes~\cite{Harvey:2014qla,Khudik:2018hkr,Blackburn:2018sfn,Blackburn:2023mlo}; as available laser intensity increases, effects related to vacuum polarisation are being added to both PIC and single-particle schemes~\cite{King:2023eeo}, and it would be interesting to test the output of the codes against the distributions above. 

The resummation methods above can also be used to compute the leading order particle spectrum; this leads to
\be
    dN_{+}(p)=\left(\frac{1}{\mathbb{b}(1)}\frac{\delta \mathbb{b}(1)}{\delta \mathcal{B}(p,p')}\Big\vert_{p'\rightarrow p}\right)\frac{\ud^3\mathbf{p}}{(2\pi)^32p_0}=\left[\mathbfbs{\mathcal{B}}\left(1-\mathbfbs{\mathcal{B}}\right)^{-1}\right]_{pp}\frac{\ud^3\mathbf{p}}{(2\pi)^32p_0} \;,
\ee
from which one immediately recovers the expected expression for the average number of pairs,
\begin{align}
    N =\sum_{n=0}^{\infty}n\mathbb{P}_{n}=\frac{\dot{\mathbb{b}}(1)}{\mathbb{b}(1)}=\textrm{Tr}\left[\mathbfbs{\mathcal{B}}\left(1-\mathbfbs{\mathcal{B}}\right)^{-1}\right] \;.
\end{align}

\subsection{Photon number at next-to-leading-order}\label{subsec:photon_no_NLO}
We have now seen how the function $\mathbb{b}(t)$, defined in (\ref{b-def}), contains information on the resummation of an infinite number of disconnected diagrams arising from the creation of pairs. Using the experience gained, we will here use $\mathbb{b}(t)$ as a generating function to explore related resummation structures in other observables, in particular the number of photons created, $\braket{N_{\gamma}}$. We begin by revisiting the leading order contribution to $\braket{N_{\gamma}}$, asking how the relatively simple expression \eqref{N-gamma-O-e} is recovered from the resummation of all contributing diagrams. The contributions to $\braket{N_{\gamma}}$ containing $\mathcal{A}_{\tadpl}$ take the form
\begin{align}\label{contrib-from-tadpole}
&\left|e^{i\mathcal{W}_{\ccd}}\right|^2\left[
\begin{tikzpicture}[baseline]
\draw[double] (-.3,.4) circle (.1);
\draw[draw=blue, smallsnake] (-.2,.4)--(0,.4);
\draw[ blue] (0,.4) --(.4,.4) ;
\draw (.55,.4) node {\small ${\bar z}_\sigma$};
\draw[dashed] (0,0.6)--(0,-0.3);
\end{tikzpicture}+
\begin{tikzpicture}[baseline]
\draw[double] (-.3,.4) circle (.1);
\draw[draw=blue, smallsnake] (-.2,.4)--(0,.4);
\draw[ blue] (0,.4) --(.4,.4) ;
\draw (.55,.4) node {\small ${\bar z}_\sigma$};
\draw[double] (0,0) circle (0.2) ;
\draw[dashed] (0,0.6)--(0,-0.3);
\end{tikzpicture}
+\frac{1}{2!}
\left(
\begin{tikzpicture}[baseline]
\draw[double] (-.3,.4) circle (.1);
\draw[draw=blue, smallsnake] (-.2,.4)--(0,.4);
\draw[ blue] (0,.4) --(.4,.4) ;
\draw (.55,.4) node {\small ${\bar z}_\sigma$};
\draw[double] (0,.15) circle (0.1) ;
\draw[double] (0,-.15) circle (0.1); 
\draw[dashed] (0,0.6)--(0,-0.3);
\end{tikzpicture}
+
\begin{tikzpicture}[baseline]
\draw[double] (-.3,.5) circle (.1);
\draw[draw=blue, smallsnake] (-.2,.5)--(0,.5);
\draw[ blue] (0,.5) --(.4,.5) ;
\draw (.55,.5) node {\small ${\bar z}_\sigma$};
\draw[double] (0,-0.3) arc (-90:90:0.3) ;
\draw[double] (0,0.3) arc (90:270:0.1) ;
\draw[double] (0,0.1) arc (90:-90:0.1) ;
\draw[double] (0,-0.1) arc (90:270:0.1); 
\draw[dashed] (0,0.6)--(0,-0.5);
\end{tikzpicture}
\right)
+ ... 
\right]+\textrm{c.c.}\\
&=\left|e^{i\mathcal{W}_{\ccd}}\right|^2\times\left[\int_{k}\mathcal{A}_{\tadpl}(k;\sigma){\bar z}_{\sigma}(k)\right]\times\left[1+ \textrm{Tr}\, \mathbfbs{\mathcal{B}}
    +\frac{1}{2!}\Big((\textrm{Tr}\, \mathbfbs{\mathcal{B}})^2+\textrm{Tr}\, (\mathbfbs{\mathcal{B}}^2)\Big)+...\right]+\textrm{c.c.}\,\\
&=\int_{k}\mathcal{A}_{\tadpl}(k;\sigma){\bar z}_{\sigma}(k)+\textrm{c.c.,}
\end{align}
where, in the last line, we used $\left|e^{i\mathcal{W}_{\ccd}}\right|^2 \mathbb{b}(1)=1$, and we recover the first term in (\ref{N-gamma-O-e}).

Contributions containing $\mathcal{A}_{\radpr}$ take the form
\begin{align}\label{contrib-from-radp}
&\left|e^{i\mathcal{W}_{\ccd}}\right|^2\left[
\begin{tikzpicture}[baseline]
\draw[smallsnake, draw=blue] (-0.2,0) arc (230:90:0.25);  
\draw[ blue] (-0.04,.45) --(.4,.45) ;
\draw (.55,.4) node {\small ${\bar z}_\sigma$};
\draw[double] (0,0) circle (0.2) ;
\draw[dashed] (0,0.6)--(0,-0.3);
\end{tikzpicture}
+\frac{1}{2!}
\Bigg(2\times
\begin{tikzpicture}[baseline]
\draw[draw=blue, smallsnake] (-.1,.22) arc (250:90:.2);
\draw[ blue] (-.03,.61) --(.4,.61) ;
\draw (.55,.6) node {\small ${\bar z}_\sigma$};
\draw[double] (0,.15) circle (0.1) ;
\draw[double] (0,-.15) circle (0.1); 
\draw[dashed] (0,0.8)--(0,-0.3);
\end{tikzpicture}
+2\times
\begin{tikzpicture}[baseline]
\draw[draw=blue, smallsnake] (-.1,.22) arc (250:90:.2);
\draw[ blue] (-.03,.61) --(.4,.61) ;
\draw (.55,.6) node {\small ${\bar z}_\sigma$};
\draw[double] (0,-0.3) arc (-90:90:0.3) ;
\draw[double] (0,0.3) arc (90:270:0.1) ;
\draw[double] (0,0.1) arc (90:-90:0.1) ;
\draw[double] (0,-0.1) arc (90:270:0.1); 
\draw[dashed] (0,0.8)--(0,-0.5);
\end{tikzpicture}
\Bigg)
+ ... 
\right]+\textrm{c.c.}\\
&=\left|e^{i\mathcal{W}_{\ccd}}\right|^2
\left[
\int_{k,p,q}{\bar z}_{\sigma}(k)\mathcal{A}_{\radpr}(p,q;k,\sigma)\frac{\delta}{\delta \mathcal{A}_{\pair}(p,q)}\left(1+ \textrm{Tr}\, \mathbfbs{\mathcal{B}}
    +\frac{1}{2!}\Big((\textrm{Tr}\, \mathbfbs{\mathcal{B}})^2+\textrm{Tr}\, (\mathbfbs{\mathcal{B}}^2)\Big)+...\right)
    \right]+\textrm{c.c.}\,\\
&=\frac{1}{\mathbb{b}(1)}\int_{k,p,q}{\bar z}_{\sigma}(k)\mathcal{A}_{\radpr}(p,q;k,\sigma)\frac{\delta \mathbb{b}(1)}{\delta \mathcal{A}_{\pair}(p,q)}=\int_{k,p,q}{\bar z}_{\sigma}(k)\textrm{Tr}\Big[
    \mathbfbs{\mathcal{A}}_{\radpr}(k;\sigma)\mathbfbs{\mathcal{A}}^{\dagger}_{\pair}(1-\mathbfbs{\mathcal{A}}_{\pair}\mathbfbs{\mathcal{A}}^{\dagger}_{\pair})^{-1} \Big]+\textrm{c.c.}
\end{align}
Adding the two contributions (\ref{contrib-from-tadpole}) and (\ref{contrib-from-radp}) recovers precisely \eqref{N-gamma-O-e}. We note briefly that the above calculation also reveals the resummation structure of the backreacted waveform $\delta z_{\sigma}$ considered earlier to be
\begin{align}\label{eq:waveform_resumm}
\delta z_{\sigma}
&=\left|e^{i\mathcal{W}_{\ccd}}\right|^2\left[
\begin{tikzpicture}[baseline]
\draw[double] (-.3,.4) circle (.1);
\draw[draw=blue, smallsnake] (-.2,.4)--(0,.4);
\draw[dashed] (0,0.6)--(0,-0.3);
\end{tikzpicture}+
\begin{tikzpicture}[baseline]
\draw[double] (-.3,.4) circle (.1);
\draw[draw=blue, smallsnake] (-.2,.4)--(0,.4);
\draw[double] (0,0) circle (0.2) ;
\draw[dashed] (0,0.6)--(0,-0.3);
\end{tikzpicture}
+\frac{1}{2!}
\left(
\begin{tikzpicture}[baseline]
\draw[double] (-.3,.4) circle (.1);
\draw[draw=blue, smallsnake] (-.2,.4)--(0,.4);
\draw[double] (0,.15) circle (0.1) ;
\draw[double] (0,-.15) circle (0.1); 
\draw[dashed] (0,0.6)--(0,-0.3);
\end{tikzpicture}
+
\begin{tikzpicture}[baseline]
\draw[double] (-.3,.5) circle (.1);
\draw[draw=blue, smallsnake] (-.2,.5)--(0,.5);
\draw[double] (0,-0.3) arc (-90:90:0.3) ;
\draw[double] (0,0.3) arc (90:270:0.1) ;
\draw[double] (0,0.1) arc (90:-90:0.1) ;
\draw[double] (0,-0.1) arc (90:270:0.1); 
\draw[dashed] (0,0.6)--(0,-0.5);
\end{tikzpicture}
\right)
+ ... \right]+\left|e^{i\mathcal{W}_{\ccd}}\right|^2\left[
\begin{tikzpicture}[baseline]
\draw[smallsnake, draw=blue] (-0.2,0) arc (230:90:0.25);  
\draw[double] (0,0) circle (0.2) ;
\draw[dashed] (0,0.6)--(0,-0.3);
\end{tikzpicture}
+\frac{1}{2!}
\Bigg(2\times
\begin{tikzpicture}[baseline]
\draw[draw=blue, smallsnake] (-.1,.22) arc (250:90:.2);
\draw[double] (0,.15) circle (0.1) ;
\draw[double] (0,-.15) circle (0.1); 
\draw[dashed] (0,0.8)--(0,-0.3);
\end{tikzpicture}
+2\times
\begin{tikzpicture}[baseline]
\draw[draw=blue, smallsnake] (-.1,.22) arc (250:90:.2);
\draw[double] (0,-0.3) arc (-90:90:0.3) ;
\draw[double] (0,0.3) arc (90:270:0.1) ;
\draw[double] (0,0.1) arc (90:-90:0.1) ;
\draw[double] (0,-0.1) arc (90:270:0.1); 
\draw[dashed] (0,0.8)--(0,-0.5);
\end{tikzpicture}
\Bigg)
+ ... 
\right]\\\label{eq:waveform_resumm_connected}
&=\begin{tikzpicture}[baseline]
\draw[double] (-.3,0) circle (.1);
\draw[draw=blue, smallsnake] (-.2,0)--(0,0);
\end{tikzpicture}+
\left(
\begin{tikzpicture}[baseline]
\draw[smallsnake, draw=blue] (-0.2,0) arc (230:90:0.25);  
\draw[double] (0,0) circle (0.2) ;
\draw[dashed] (0,0.6)--(0,-0.3);
\end{tikzpicture}+
\begin{tikzpicture}[baseline]
\draw[draw=blue, smallsnake] (-.1,.22) arc (250:90:.2);
\draw[double] (0,-0.3) arc (-90:90:0.3) ;
\draw[double] (0,0.3) arc (90:270:0.1) ;
\draw[double] (0,0.1) arc (90:-90:0.1) ;
\draw[double] (0,-0.1) arc (90:270:0.1); 
\draw[dashed] (0,0.8)--(0,-0.5);
\end{tikzpicture}+
...
\right)\\
&=\mathcal{A}_{\tadpl}(k;\sigma)
    +
    \textrm{Tr}\Big[
    \mathbfbs{\mathcal{A}}_{\radpr}(k;\sigma)\mathbfbs{\mathcal{A}}^{\dagger}_{\pair}(1-\mathbfbs{\mathcal{A}}_{\pair}\mathbfbs{\mathcal{A}}^{\dagger}_{\pair})^{-1} \Big] \;.
\end{align}
Observe that to arrive at the second line of the above we used~\eqref{b-def} to cancel the normalisation factor against the infinite series of terms with non-radiating horse-shoes. This emphasises the subtle diagrammatic cancellations which occur in order to, essentially, convert from in-out to in-in quantities.

Notice how we were able, here at LO, to use $\mathbb{b}(1)$ as a generating function from which to derive the contribution from all disconnected diagrams. The same approach leads, at next-to-leading order, to the formal expression
\begin{align}
    \braket{N_{\gamma}}&=\int_k  {\bar z}_\sigma(k)z_\sigma(k) + \Big[{\bar z}_\sigma(k)\delta z_\sigma(k) + z_\sigma(k)\delta {\bar z}_\sigma(k)\Big]\\\nonumber
   &+\int_k  \bar{\mathcal{A}}_{\tadpl}(k;\sigma)\mathcal{A}_{\tadpl}(k;\sigma)+2\int_{k,p,q}\textrm{Re}\left[
   \bar{\mathcal{A}}_{\tadpl}(k;\sigma)\mathcal{A}_{\radpr}(p,q,k;\sigma)\frac{1}{\mathbb{b}(1)}\frac{\delta\mathbb{b}(1)}{\delta\mathcal{A}_{\pair}(p,q)}\right]\\\nonumber
   &+\int_{p_1,q_1,p_2,q_2,k}\mathcal{A}_{\radpr}(p_1,q_1,k;\sigma)\bar{\mathcal{A}}_{\radpr}(p_2,q_2,k;\sigma)\frac{1}{\mathbb{b}(1)}\frac{\delta^2\mathbb{b}(1)}{\delta\mathcal{A}_{\pair}(p_1,q_1)\delta\bar{\mathcal{A}}_{\pair}(p_2,q_2)}+\mathcal{O}_{F}(e^3)
\end{align}
This expression simplifies after performing the functional derivatives. Dropping the $\mathcal{O}_F(e)$ term already studied and focussing on the $\mathcal{O}_F(e^2)$ contribution only,
\begin{align}\label{eq:N_gamma_o2_part}
  \braket{N_{\gamma}}\bigg|_{e^2} &=\int_k \delta z_\sigma(k)\delta {\bar z}_\sigma(k)\\\nonumber
  &+\int_{k}\textrm{Tr}\left[(1-\mathbfbs{\mathcal{A}_{\pair}}\mathbfbs{\mathcal{A}^{\dagger}_{\pair}})^{-1}\mathbfbs{\mathcal{A}_{\pair}}\mathbfbs{\mathcal{A}^{\dagger}_{\radpr}}(k;\sigma)(1-\mathbfbs{\mathcal{A}_{\pair}}\mathbfbs{\mathcal{A}^{\dagger}_{\pair}})^{-1}\mathbfbs{\mathcal{A}_{\radpr}}(k;\sigma)\mathbfbs{\mathcal{A}^{\dagger}_{\pair}}\right]\\\nonumber
    &+\int_{k}\textrm{Tr}\left[(1-\mathbfbs{\mathcal{A}_{\pair}}\mathbfbs{\mathcal{A}^{\dagger}_{\pair}})^{-1}\mathbfbs{\mathcal{A}_{\radpr}}(k;\sigma)\mathbfbs{\mathcal{A}^{\dagger}_{\radpr}}(k;\sigma)\right] \;.
\end{align}
It is worth analysing the different terms in \eqref{eq:N_gamma_o2_part} in terms of the topology of contributing diagrams. The first term arises from diagrams where the cut-photon line runs between two otherwise connected parts. That is, symbolically,
\begin{align}
 \int_k \delta z_\sigma(k)\delta {\bar z}_\sigma(k)\supset \begin{tikzpicture}[baseline]
 \draw[draw=lightgray, fill=lightgray] (-.3,.3) rectangle ++(.6,.6) ;
 \draw[draw=lightgray, fill=lightgray] (.3,-.3) rectangle ++(-.6,-.6) ;
 \draw[smallsnake,blue] (0,0) arc (90:200:.3);
 \draw[smallsnake,blue] (0,0) arc (-90:20:.3);
 \draw[dashed] (0,1)--(0,-1);
 \end{tikzpicture}\,
\end{align}
where each of the shaded boxes represents, as evident from \eqref{eq:waveform_resumm_connected}, a connected contribution containing only particles / antiparticles. The second and third lines in \eqref{eq:N_gamma_o2_part}, on the other hand, correspond to diagrams in which the cut photon line shown runs between two parts of an already connected diagram, this again being indicated by a shaded box. Symbolically, this \textit{connected} contribution to $\braket{N_{\gamma}}\big|_{e^2}$, indicated by a subscript `ctd.', may be illustrated by
\begin{align}
 \left. \begin{matrix} \braket{N_{\gamma}}\bigg|_{e^2,\text{ ctd.}}\equiv 
\int_{k}\textrm{Tr}\left[(1-\mathbfbs{\mathcal{A}_{\pair}}\mathbfbs{\mathcal{A}^{\dagger}_{\pair}})^{-1}\mathbfbs{\mathcal{A}_{\pair}}\mathbfbs{\mathcal{A}^{\dagger}_{\radpr}}(k;\sigma)(1-\mathbfbs{\mathcal{A}_{\pair}}\mathbfbs{\mathcal{A}^{\dagger}_{\pair}})^{-1}\mathbfbs{\mathcal{A}_{\radpr}}(k;\sigma)\mathbfbs{\mathcal{A}^{\dagger}_{\pair}}\right]\\
+\,\,\int_{k}\textrm{Tr}\left[(1-\mathbfbs{\mathcal{A}_{\pair}}\mathbfbs{\mathcal{A}^{\dagger}_{\pair}})^{-1}\mathbfbs{\mathcal{A}_{\radpr}}(k;\sigma)\mathbfbs{\mathcal{A}^{\dagger}_{\radpr}}(k;\sigma)\right]
\end{matrix}\right\} \supset \begin{tikzpicture}[baseline]
\draw[draw=lightgray, fill=lightgray] (-.25,1.2) rectangle ++(.5,-2.4);
\draw[double,draw opacity=.2] (0,0.75) arc (90:270:.25) ;
\draw[double,draw opacity=.2] (0,-0.85) arc (-90:90:.25) ;
\draw[dashed] (0,1.4)--(0,-1.5);
\draw[draw=blue, smallsnake] (-.25,.5) to [bend right =60] (0,0);
\draw[draw=blue, smallsnake] (.25,-.6) to [bend right =60] (0,0);
 \end{tikzpicture}\,,
\end{align}
where, again, the shaded region represents an arbitrary connected term. Adding the LO and NLO contributions, the number of photons takes the form
\begin{align}\label{N-gamma-2-final}
    \braket{N_{\gamma}}&=\int_k\left|z_{\sigma}(k)+\delta z_{\sigma}(k)\right|^2+ \braket{N_{\gamma}}\bigg|_{e^2,\text{ ctd.}}+\mathcal{O}_F(e^4)
\end{align}
This expression reinforces the result, c.f.~the discussion under (\ref{33}), that the final state is not a coherent state -- the number density (\ref{N-gamma-2-final}) is not that of a coherent state, due to the presence of the connected contribution.
\subsection{Vacuum persistence at the next-to-leading order}\label{sec:NLO_vac_to_vac}
%
We turn to the calculation of NLO contributions to the vacuum persistence amplitude. This brings together much of the work above, because the NLO terms contain both loop and backreaction effects, i.e.~corrections arising from off-shell and on-shell photons, respectively. We will see that the final correction to vacuum persistence emerges from an interesting interplay of these two types of contribution.

Continuing with the approach taken above, we again exploit the function $\mathbb{b}(t)$ to generate the NLO corrections, which start at $\mathcal{O}_{F}(e^2)$. To this end we start with the expression for $\mathbb{b}(t)$ given in \eqref{eq:b(t)_is_Tr_exp} and rewrite it as
\begin{align}
   \mathbb{b}(t)=\exp\bigg(\sum_{j=1}^{\infty}t^j\frac{\lambda_j}{j}\bigg)\,, 
\end{align}
where, as in \eqref{eq:TrBJ_is_lambdaj}, we denote $\lambda_j=\textrm{Tr}(\mathbfbs{\mathcal{B}}^j)$, which corresponds to a connected cut-diagram of $j$ pairs. Corrections to $\mathbb{b}(t)$ start at order $\mathcal{O}_{F}(e^2)$, so if we write $\mathbb{b}^{(2)}(t)$ for this corrected generator, we may parametrise it as
\begin{align}
    \mathbb{b}^{(2)}(t)=\mathbb{b}(t)+\Delta^{(2)}\mathbb{b}(t)=\mathbb{b}(t)\exp\left[\frac{\Delta^{(2)}\mathbb{b}(t)}{\mathbb{b}(t)}\right]+\mathcal{O}_{F}(e^4)
\end{align}
It is therefore natural to study the corrections to the log of $\mathbb{b}(t)$ which, in turn, can be studied by looking at the corrections to the relevant connected-diagrams $\lambda_{j}$. Specifically, we can define the $\mathcal{O}_{F}(e^2)$ analogue of $\mathbb{b}(t)$ as
\begin{align}\label{eq:O_2_b_t}
   \mathbb{b}^{(2)}(t)\equiv\exp\left[\int_k \mathcal{A}_{\tadpl}(k;\sigma)
   \bar{\mathcal{A}}_{\tadpl}(k;\sigma)   \right]\exp\left(\sum_{j=1}^{\infty}t^j\frac{\lambda^{(2)}_j}{j}\right)+\mathcal{O}_F(e^4)\quad;\quad \lambda^{(2)}_j\equiv \lambda_j +\Delta^{(2)}\lambda_j \,,
\end{align}
where $\lambda^{(2)}_j$ is the $\mathcal{O}_{F}(e^2)$ analogue of $\textrm{Tr}(\mathbfbs{\mathcal{B}}^j)$ and $\Delta^{(2)}\lambda_j$ is the corresponding  correction. Before discussing these corrections for arbitrary $j$, it is helpful to examine the structure in the simplest cases of $j=1$ and $j=2$.  Firstly,
\begin{align}\label{eq:delta_lambda_1}
 \Delta^{(2)}\lambda_1=\left( 
 \begin{tikzpicture}[baseline]
\draw[double] (0,0) circle (.5);
\draw[smallsnake, blue] (-.5,0) -- (-.8,0);
\draw[double] (-1,0) circle (.2);
\draw[dashed] (0,-1.2) -- (0,1.2);
\end{tikzpicture}
+
\begin{tikzpicture}[baseline]
\draw[double] (0,0) circle (.5);
\draw[smallsnake, blue] (-.353,.353) arc (225:25:.4);
\draw[double] (.41,.67) circle (.2);
\draw[dashed] (0,-1.2) -- (0,1.2);
\end{tikzpicture}
+
\begin{tikzpicture}[baseline]
\draw[double] (0,0) circle (.5);
\draw[smallsnake, blue] (-.353,.353) arc (45:-45:.5);
\draw[dashed] (0,-1.2) -- (0,1.2);
\end{tikzpicture}
+
\textrm{c.c.}\right)\quad+\quad
\begin{tikzpicture}[baseline]
\draw[double] (0,0) circle (.5);
\draw[smallsnake, blue] (-.5,0) -- (.5,0);
\draw[dashed] (0,-1.2) -- (0,1.2);
\end{tikzpicture} \;.
\end{align}
Note the contribution to $\Delta^{(2)}\lambda_1$ from the diagrams in the large brackets above can be generated by replacing $\mathcal{A}_{\pair}$ with the loop-corrected quantity
\begin{align}
    \tilde{\mathcal{A}}^{(2)}_{\pair}(p,q)&:=\mathcal{A}^{(2)}_{\pair}(p,q)+\int_k\mathcal{A}_{\radpr}(p,q;k,\sigma)
    \bar{\mathcal{A}}_{\tadpl}(k;\sigma)\quad,\\
\textrm{where}\quad\mathcal{A}^{(2)}_{\pair}&=
\begin{tikzpicture}[baseline]
 \draw[double] (0,.3) -- (.3,.3);
  \draw[double] (0,-.3) -- (.3,-.3);
\draw[double] (0,0.3) arc (90:270:.3);
\end{tikzpicture}+ 
\begin{tikzpicture}[baseline]
 \draw[double] (0,.3) -- (.3,.3);
  \draw[double] (0,-.3) -- (.3,-.3);
\draw[double] (0,0.3) arc (90:270:.3);
\draw[draw=blue, smallsnake] (-.21,.21) arc (45:-45:.3) ;
\end{tikzpicture}+
\begin{tikzpicture}[baseline]
 \draw[double] (0,.3) -- (.3,.3);
  \draw[double] (0,-.3) -- (.3,-.3);
\draw[double] (0,0.3) arc (90:270:.3);
\draw[draw=blue, smallsnake] (-.3,0)--(-.5,0);
\draw[double] (-.6,0) circle (.1);
\end{tikzpicture}
\end{align}
In contrast, the final term in \eqref{eq:delta_lambda_1}, is quadratic in the variables $(\mathcal{A}_{\radpr},\bar{\mathcal{A}}_{\radpr})$.

For $j>1$, we will find, as at $j=1$, a subset of diagrams that may be accounted for by replacing $\mathcal{A}_{\pair}$ with $\mathcal{A}^{(2)}_{\pair}$, but other types of contribution also start to appear, which cannot be expressed using only $\mathcal{A}^{(2)}_{\pair}$. A representative contribution to $\Delta^{(2)}\lambda_2$ is
\begin{align}
    \Delta^{(2)}\lambda_2 \supset
 \left(
 \begin{tikzpicture}[baseline]
\draw[double] (0,.9) arc (90:270:.3);
\draw[double] (0,.3) arc (90:-90:.3); 
\draw[double] (0,-.3) arc (90:270:.3) ; 
\draw[double] (0,-.9) arc (-90:90:.9); 
\draw[dashed] (0,1.2)--(0,-1.2);
\draw[smallsnake,blue] (-.3,.6) arc (135:-45:.4);
\end{tikzpicture}+
\begin{tikzpicture}[baseline]
\draw[double] (0,.6) circle (.3);
\draw[double] (0,-.6) circle (.3);  
\draw[dashed] (0,1.2)--(0,-1.2);
\draw[smallsnake,blue] (-.3,.6) arc (135:275:.36);
\draw[smallsnake,blue] (.3,-.6) arc (-45:95:.36);
\end{tikzpicture}+
\begin{tikzpicture}[baseline]
\draw[double] (0,.9) arc (90:270:.3);
\draw[double] (0,.3) arc (90:-90:.3); 
\draw[double] (0,-.3) arc (90:270:.3) ; 
\draw[double] (0,-.9) arc (-90:90:.9); 
\draw[dashed] (0,1.2)--(0,-1.2);
\draw[smallsnake,blue] (-.3,.6) arc (135:225:.8);
\end{tikzpicture}+
\begin{tikzpicture}[baseline]
\draw[double] (0,.6) circle (.3);
\draw[double] (0,-.6) circle (.3);  
\draw[dashed] (0,1.2)--(0,-1.2);
\draw[smallsnake,blue] (-.3,.6) arc (135:225:.8);
\end{tikzpicture}+
\textrm{c.c.}\right)   
\end{align}
While the first two terms of the above are analogous to the final (`quadratic') diagram in~\eqref{eq:delta_lambda_1}, the final two terms arise from the $\mathcal{O}_F(e^2)$ correction to the amplitude for creation of two pairs. To account for the effect of this contribution it is evidently convenient to define
\begin{align}
    \mathcal{A}_{\pairsqr}(p_1,q_1;p_2,q_2)=
\begin{tikzpicture}[baseline]
 \draw[double] (0,.75) -- (.3,.75);
  \draw[double] (0,.15) -- (.3,.15);
\draw[double] (0,.75) arc (90:270:.3);
\draw[double] (0,-.75) -- (.3,-.75);
  \draw[double] (0,-.15) -- (.3,-.15);
\draw[double] (0,-.15) arc (90:270:.3);
\draw[draw=blue, smallsnake] (-.3,.3) to [bend right =60] (-.3,-.3);
\draw (.55,.75) node {$p_1$};
\draw (.55,.15) node {$q_1$};
\draw (.55,-.75) node {$q_2$};
\draw (.55,-.15) node {$p_2$};
\end{tikzpicture}+
\begin{tikzpicture}[baseline]
 \draw[double] (0,.75) -- (.3,.75);
  \draw[double] (0,.15) -- (.3,.15);
\draw[double] (0,.75) arc (90:270:.3);
\draw[double] (0,-.75) -- (.3,-.75);
  \draw[double] (0,-.15) -- (.3,-.15);
\draw[double] (0,-.15) arc (90:270:.3);
\draw[draw=blue, smallsnake] (-.3,.3) to [bend right =60] (-.3,-.3);
\draw (.55,.75) node {$p_2$};
\draw (.55,.15) node {$q_1$};
\draw (.55,-.75) node {$q_2$};
\draw (.55,-.15) node {$p_1$};
\end{tikzpicture} \;,
\end{align}
which contains contributions from both the Coulomb interaction between pairs and from a cascade process; these effects can formally be distinguished by considering, respectively, the potential and causal parts of the photon propagator. By analysing the cases $j>2$ in a similar way one finds that $\mathcal{A}^{(2)}_{\pair}$, $\mathcal{A}_{\radpr}$, $\bar{\mathcal{A}}_{\radpr}$ and $\mathcal{A}_{\pairsqr}$ are all that one requires to study~$\lambda^{(2)}_j$.

Generalizing the discussion above, we can write
\begin{align}\label{eq:O_2_lambda_j}
    \lambda^{(2)}_j&=\textrm{Tr}[(\mathbfbs{\mathcal{A}}^{(2)}_{\pair}\mathbfbs{\mathcal{A}}^{(2)\dagger}_{\pair})^j]\\\nonumber
    &+j\,\mathbfbs{[}t^j\mathbfbs{]}_{\rm coeff.}\left\{2\int_{k,p,q}\textrm{Re}\left[
    \bar{\mathcal{A}}_{\tadpl}(k;\sigma)\mathcal{A}_{\radpr}(p,q,k;\sigma)\frac{1}{\mathbb{b}(t)}\frac{\delta\mathbb{b}(t)}{\delta\mathcal{A}_{\pair}(p,q)}\right]\right.\\\nonumber
    &+\left.\int_{p_1,q_1,p_2,q_2,k}\mathcal{A}_{\radpr}(p_1,q_1,k;\sigma)
    \bar{\mathcal{A}}_{\radpr}(p_2,q_2,k;\sigma)\frac{1}{\mathbb{b}(t)}\frac{\delta^2\mathbb{b}(t)}{\delta\mathcal{A}_{\pair}(p_1,q_1)
    \delta\bar{\mathcal{A}}_{\pair}(p_2,q_2)}\right\}\\\nonumber
    &+\frac{1}{2!}\left(\int_{p_1,p_2,q_1,q_2}\mathcal{A}_{\pairsqr}(p_1,q_1;p_2,q_2)\frac{\delta^2}{\delta\mathcal{A}_{\pair}(p_1,q_1)\delta\mathcal{A}_{\pair}(p_2,q_2)}\textrm{Tr}[(\mathbfbs{\mathcal{A}}_{\pair}\mathbfbs{\mathcal{A}}^{\dagger}_{\pair})^j]+\textrm{c.c.}\right)+\mathcal{O}_F(e^4)\,,
\end{align}
where $\mathbfbs{[}t^j\mathbfbs{]}_{\rm coeff.}\left\{x(t)\right\}$ is shorthand for the coefficient of $t^j$ in the Taylor series of $x(t)$. The first, second, and third lines of (\ref{eq:O_2_lambda_j}) correspond, respectively, to the loop-corrected amplitude for creation of a single pair, the radiation of a photon by a created pair, and the photon-mediated interaction between two created pairs. The interplay of these contributions manifests in the most interesting form in the expression for the $\mathcal{O}_{F}(e^2)$ (log of) the vacuum-persistence probability, which, from \eqref{eq:O_2_b_t}, takes the form
\begin{align}
    \log(\mathbb{P}^{(2)})\equiv -\sum_{j=1}^{\infty}\frac{\lambda^{(2)}_j}{j}-\int_k \mathcal{A}_{\tadpl}(k;\sigma)
    \bar{\mathcal{A}}_{\tadpl}(k;\sigma)
\end{align}
Using \eqref{eq:O_2_lambda_j} and the discussion surrounding $\braket{N_{\gamma}}\vert_{e^2}$, we find that
\begin{align}\label{eq:log_P_O_2}
    &\log(\mathbb{P}^{(2)})    \\\nonumber
    &=\log\left[\mathbb{P}\left(\mathcal{A}_{\pair}\rightarrow \mathcal{A}^{(2)}_{\pair}\right)\right]-\frac{1}{2!}\left(\int_{p_1,p_2,q_1,q_2}\mathcal{A}_{\pairsqr}(p_1,q_1;p_2,q_2)\frac{\delta^2}{\delta\mathcal{A}_{\pair}(p_1,q_1)\delta\mathcal{A}_{\pair}(p_2,q_2)}\log\mathbb{P}+\textrm{c.c.}\right)-\braket{N_{\gamma}}\Bigg\vert_{e^2}+\mathcal{O}_{F}(e^4)\,,
\end{align}
where the functional derivative above can be performed using \eqref{eq:b(t)_is_Tr_exp} for $\mathbb{P}=\mathbb{b}^{-1}(1)$, to transform the same into a series expansion. Paralleling the discussion of Sec.~\ref{subsec:vp1}, we can now write the probability for creation of $n$ pairs at~$\mathcal{O}_{F}(e^2)$~as
\begin{align}\label{eq:Bell_distribution_O2}
  \mathbb{P}^{(2)}_{n}=  \frac{\exp\left(-\sum_{j=1}^{\infty}t^j\frac{\lambda^{(2)}_j}{j}\right)}{n!}B_{n}(\lambda^{(2)}_1,\lambda^{(2)}_2,2!\lambda^{(2)}_3,...,(n-1)!\lambda^{(2)}_n)\,,
\end{align}
which is analogous to \eqref{eq:Bell_distribution}. This implies that the average number of pairs is given by
\begin{align}
    N^{(2)}=\sum_{n=1}^{\infty}n\mathbb{P}^{(2)}_{n}=\sum_{n=1}^{\infty}\lambda^{(2)}_j
\end{align}
Of course, the right-hand side of the above can also be rearranged into a form analogous to \eqref{eq:log_P_O_2}, showing contributions from different physical effects. Additionally, if we consider the case where there is a hierarchy of the form $\lambda^{(2)}_{j+1}\ll\lambda^{(2)}_{j}\ll \lambda^{(2)}_{1}$, as we did in \ref{subsub:pair_dist}, the statistics of pair creation process is described by a deformed Poisson distribution. In particular, if we restrict to $\mathcal{O}(\lambda^{(2)}_{2})$, then we find the probabilities to be essentially of the form in~\eqref{Pn-norm}, but with $\lambda_1$ and $\lambda_2$ replaced by, respectively, $\lambda^{(2)}_1$ and $\lambda^{(2)}_2$. Physically, the former replacement accounts for the 2-loop effects as well as the radiation of a single photon. The effects of interaction between pairs (including the Coulumb force as well as lowest order cascade process), however, start to first appear in $\lambda^{(2)}_2$. In this sense, it is \textit{necessary} to consider the deformed Poisson distribution in \eqref{Pn-norm}, with the parameters adjusted in the manner we just described, to model pair creation statistics that account for all physical effects at $\mathcal{O}_F(e^2)$.

\section{Conclusions}\label{sec:concs}
%
We have discussed a systematic approach to backreaction in asymptotic observables, in strong field (scalar) QED. An advantage of our approach is that it expresses observables in terms of familiar amplitudes calculated on a fixed background. In particular, one can compute expectation values of operators systematically, in the Furry expansion, and so leverage the vast literature on amplitudes in fixed backgrounds.

We began by considering a coherent state of photons, modelling a (strong) field, and discussed the form of the time-evolved state and how it encodes information on various physical phenomena, driven by the field. We focused mainly on three observables: the number of created pairs, the number of created photons, and the waveform. These observables can be expressed as a series of disconnected background-Feynman diagrams at a given order in the Furry expansion. We have illustrated how these series can be exactly resummed. The effects of backreaction appear already in the leading order expressions for the mean number of emitted photons and the waveform, at next-to-leading order in the mean number of pairs.

As an example, we applied our formalism to study backreaction in the case of a delta-function electric field. This is a useful toy model that allows several exact computations. For instance, we were able to compute the leading fall-off behaviour of the back-reacted electromagnetic field strength. As expected from the symmetry of the background, there is no contribution to the waveform, at leading order, at future null infinity. However, at time-like future infinity there is a power-law damped oscillatory electromagnetic field sustained by the current generated by the created pairs. 

In another example, we studied not just the mean number of pairs produced (in an arbitrary field) but also the statistics of the pair creation process. In particular, an analysis of disconnected diagrams and their resummation revealed the probability distribution for the creation of $n$ pairs, which we showed was expressible in terms of Bell's polynomials. This distribution reduces, in the weak-field limit, to Poissonian, but to properly account for all $\mathcal{O}_F(e^2)$ effects in pair creation we showed that one must consider at least a one-parameter deformation of Poisson.

The number of physical effects contributing to observables increases rapidly at higher orders in the Furry expansion, which we illustrated using the vacuum persistence probability at $\mathcal{O}_F(e^2)$. By once again resumming the relevant disconnected diagrams we found an interesting interplay of two-loop effects, radiation of photons, the Coulomb interaction between pairs, and cascade processes. We also showed how the vacuum persistence probability can be arranged to clearly exhibit these different contributions.

It would be interesting to see how this formalism can be applied to situations in which the initial coherent state is chosen to model laser-laser collisions. It would be also interesting to extend our results to initial states containing several particles and/or photons, in addition to the coherent state. Such an extension would be relevant to, for example, to laser-particle collisions. We plan to explore these avenues in the future.

For simplicity we discussed only scalar QED. However, most of our analysis can be straightforwardly extended to QED proper, with certain obvious departures. Fermi-Dirac statistics implies that many of the cut diagrams appearing in our results acquire an appropriate power of `$-1$'. This can, in most instances, be effected by the replacement $t\rightarrow -t$ in the generating function $\mathbb{b}(t)$. The Bogoliubov constraints~\eqref{Bogol-def} should also be modified analogously. In QED proper one would also have an extra index to account for the particle spin. This would be carried along with the `matrix multiplications' introduced in \eqref{matrix-sewing-def}. Therefore, at least for the observables considered here, the extension to spinor QED does not pose any real challenge.  It should in any case be clear that the essentials of our diagrammatic arguments carry over.

\begin{acknowledgments}
A.I.~thanks Rafa Aoude and Donal O'Connell for useful discussions and for sharing a draft of~\cite{Aoude:2024sve}. The authors are supported by the EPSRC Standard Grants  EP/X02413X/1 (PC, JPE) and EP/X024199/1 (AI, KR), and the STFC consolidator grant ``Particle Theory at the Higgs Centre,'' ST/X000494/1 (AI). 	
\end{acknowledgments}

\bibliography{BR_SQED}

\begin{thebibliography}{84}%
\makeatletter
\providecommand \@ifxundefined [1]{%
 \@ifx{#1\undefined}
}%
\providecommand \@ifnum [1]{%
 \ifnum #1\expandafter \@firstoftwo
 \else \expandafter \@secondoftwo
 \fi
}%
\providecommand \@ifx [1]{%
 \ifx #1\expandafter \@firstoftwo
 \else \expandafter \@secondoftwo
 \fi
}%
\providecommand \natexlab [1]{#1}%
\providecommand \enquote  [1]{``#1''}%
\providecommand \bibnamefont  [1]{#1}%
\providecommand \bibfnamefont [1]{#1}%
\providecommand \citenamefont [1]{#1}%
\providecommand \href@noop [0]{\@secondoftwo}%
\providecommand \href [0]{\begingroup \@sanitize@url \@href}%
\providecommand \@href[1]{\@@startlink{#1}\@@href}%
\providecommand \@@href[1]{\endgroup#1\@@endlink}%
\providecommand \@sanitize@url [0]{\catcode `\\12\catcode `\$12\catcode `\&12\catcode `\#12\catcode `\^12\catcode `\_12\catcode `\%12\relax}%
\providecommand \@@startlink[1]{}%
\providecommand \@@endlink[0]{}%
\providecommand \url  [0]{\begingroup\@sanitize@url \@url }%
\providecommand \@url [1]{\endgroup\@href {#1}{\urlprefix }}%
\providecommand \urlprefix  [0]{URL }%
\providecommand \Eprint [0]{\href }%
\providecommand \doibase [0]{http://dx.doi.org/}%
\providecommand \selectlanguage [0]{\@gobble}%
\providecommand \bibinfo  [0]{\@secondoftwo}%
\providecommand \bibfield  [0]{\@secondoftwo}%
\providecommand \translation [1]{[#1]}%
\providecommand \BibitemOpen [0]{}%
\providecommand \bibitemStop [0]{}%
\providecommand \bibitemNoStop [0]{.\EOS\space}%
\providecommand \EOS [0]{\spacefactor3000\relax}%
\providecommand \BibitemShut  [1]{\csname bibitem#1\endcsname}%
\let\auto@bib@innerbib\@empty
\bibitem [{\citenamefont {Cole}\ \emph {et~al.}(2018)\citenamefont {Cole} \emph {et~al.}}]{Cole:2017zca}%
  \BibitemOpen
  \bibfield  {author} {\bibinfo {author} {\bibfnamefont {J.~M.}\ \bibnamefont {Cole}} \emph {et~al.},\ }\href {\doibase 10.1103/PhysRevX.8.011020} {\bibfield  {journal} {\bibinfo  {journal} {Phys. Rev. X}\ }\textbf {\bibinfo {volume} {8}},\ \bibinfo {pages} {011020} (\bibinfo {year} {2018})},\ \Eprint {http://arxiv.org/abs/1707.06821} {arXiv:1707.06821 [physics.plasm-ph]} \BibitemShut {NoStop}%
\bibitem [{\citenamefont {Poder}\ \emph {et~al.}(2018)\citenamefont {Poder} \emph {et~al.}}]{Poder:2017dpw}%
  \BibitemOpen
  \bibfield  {author} {\bibinfo {author} {\bibfnamefont {K.}~\bibnamefont {Poder}} \emph {et~al.},\ }\href {\doibase 10.1103/PhysRevX.8.031004} {\bibfield  {journal} {\bibinfo  {journal} {Phys. Rev. X}\ }\textbf {\bibinfo {volume} {8}},\ \bibinfo {pages} {031004} (\bibinfo {year} {2018})},\ \Eprint {http://arxiv.org/abs/1709.01861} {arXiv:1709.01861 [physics.plasm-ph]} \BibitemShut {NoStop}%
\bibitem [{\citenamefont {Abramowicz}\ \emph {et~al.}(2021)\citenamefont {Abramowicz} \emph {et~al.}}]{Abramowicz:2021zja}%
  \BibitemOpen
  \bibfield  {author} {\bibinfo {author} {\bibfnamefont {H.}~\bibnamefont {Abramowicz}} \emph {et~al.},\ }\href {\doibase 10.1140/epjs/s11734-021-00249-z} {\bibfield  {journal} {\bibinfo  {journal} {Eur. Phys. J. ST}\ }\textbf {\bibinfo {volume} {230}},\ \bibinfo {pages} {2445} (\bibinfo {year} {2021})},\ \Eprint {http://arxiv.org/abs/2102.02032} {arXiv:2102.02032 [hep-ex]} \BibitemShut {NoStop}%
\bibitem [{\citenamefont {Ahmadiniaz}\ \emph {et~al.}(2024)\citenamefont {Ahmadiniaz} \emph {et~al.}}]{Ahmadiniaz:2024xob}%
  \BibitemOpen
  \bibfield  {author} {\bibinfo {author} {\bibfnamefont {N.}~\bibnamefont {Ahmadiniaz}} \emph {et~al.},\ }\href@noop {} {\  (\bibinfo {year} {2024})},\ \Eprint {http://arxiv.org/abs/2405.18063} {arXiv:2405.18063 [physics.ins-det]} \BibitemShut {NoStop}%
\bibitem [{\citenamefont {Los}\ \emph {et~al.}(2024)\citenamefont {Los} \emph {et~al.}}]{Los:2024ysw}%
  \BibitemOpen
  \bibfield  {author} {\bibinfo {author} {\bibfnamefont {E.~E.}\ \bibnamefont {Los}} \emph {et~al.},\ }\href@noop {} {\  (\bibinfo {year} {2024})},\ \Eprint {http://arxiv.org/abs/2407.12071} {arXiv:2407.12071 [hep-ph]} \BibitemShut {NoStop}%
\bibitem [{\citenamefont {Furry}(1951)}]{Furry:1951bef}%
  \BibitemOpen
  \bibfield  {author} {\bibinfo {author} {\bibfnamefont {W.~H.}\ \bibnamefont {Furry}},\ }\href {\doibase 10.1103/PhysRev.81.115} {\bibfield  {journal} {\bibinfo  {journal} {Phys. Rev.}\ }\textbf {\bibinfo {volume} {81}},\ \bibinfo {pages} {115} (\bibinfo {year} {1951})}\BibitemShut {NoStop}%
\bibitem [{\citenamefont {Gonoskov}\ \emph {et~al.}(2022)\citenamefont {Gonoskov}, \citenamefont {Blackburn}, \citenamefont {Marklund},\ and\ \citenamefont {Bulanov}}]{Gonoskov:2021hwf}%
  \BibitemOpen
  \bibfield  {author} {\bibinfo {author} {\bibfnamefont {A.}~\bibnamefont {Gonoskov}}, \bibinfo {author} {\bibfnamefont {T.~G.}\ \bibnamefont {Blackburn}}, \bibinfo {author} {\bibfnamefont {M.}~\bibnamefont {Marklund}}, \ and\ \bibinfo {author} {\bibfnamefont {S.~S.}\ \bibnamefont {Bulanov}},\ }\href {\doibase 10.1103/RevModPhys.94.045001} {\bibfield  {journal} {\bibinfo  {journal} {Rev. Mod. Phys.}\ }\textbf {\bibinfo {volume} {94}},\ \bibinfo {pages} {045001} (\bibinfo {year} {2022})},\ \Eprint {http://arxiv.org/abs/2107.02161} {arXiv:2107.02161 [physics.plasm-ph]} \BibitemShut {NoStop}%
\bibitem [{\citenamefont {Fedotov}\ \emph {et~al.}(2023)\citenamefont {Fedotov}, \citenamefont {Ilderton}, \citenamefont {Karbstein}, \citenamefont {King}, \citenamefont {Seipt}, \citenamefont {Taya},\ and\ \citenamefont {Torgrimsson}}]{Fedotov:2022ely}%
  \BibitemOpen
  \bibfield  {author} {\bibinfo {author} {\bibfnamefont {A.}~\bibnamefont {Fedotov}}, \bibinfo {author} {\bibfnamefont {A.}~\bibnamefont {Ilderton}}, \bibinfo {author} {\bibfnamefont {F.}~\bibnamefont {Karbstein}}, \bibinfo {author} {\bibfnamefont {B.}~\bibnamefont {King}}, \bibinfo {author} {\bibfnamefont {D.}~\bibnamefont {Seipt}}, \bibinfo {author} {\bibfnamefont {H.}~\bibnamefont {Taya}}, \ and\ \bibinfo {author} {\bibfnamefont {G.}~\bibnamefont {Torgrimsson}},\ }\href {\doibase 10.1016/j.physrep.2023.01.003} {\bibfield  {journal} {\bibinfo  {journal} {Phys. Rept.}\ }\textbf {\bibinfo {volume} {1010}},\ \bibinfo {pages} {1} (\bibinfo {year} {2023})},\ \Eprint {http://arxiv.org/abs/2203.00019} {arXiv:2203.00019 [hep-ph]} \BibitemShut {NoStop}%
\bibitem [{\citenamefont {Seipt}\ \emph {et~al.}(2017)\citenamefont {Seipt}, \citenamefont {Heinzl}, \citenamefont {Marklund},\ and\ \citenamefont {Bulanov}}]{Seipt:2016fyu}%
  \BibitemOpen
  \bibfield  {author} {\bibinfo {author} {\bibfnamefont {D.}~\bibnamefont {Seipt}}, \bibinfo {author} {\bibfnamefont {T.}~\bibnamefont {Heinzl}}, \bibinfo {author} {\bibfnamefont {M.}~\bibnamefont {Marklund}}, \ and\ \bibinfo {author} {\bibfnamefont {S.~S.}\ \bibnamefont {Bulanov}},\ }\href {\doibase 10.1103/PhysRevLett.118.154803} {\bibfield  {journal} {\bibinfo  {journal} {Phys. Rev. Lett.}\ }\textbf {\bibinfo {volume} {118}},\ \bibinfo {pages} {154803} (\bibinfo {year} {2017})},\ \Eprint {http://arxiv.org/abs/1605.00633} {arXiv:1605.00633 [hep-ph]} \BibitemShut {NoStop}%
\bibitem [{\citenamefont {Bell}\ and\ \citenamefont {Kirk}(2008)}]{BellKirk:2008}%
  \BibitemOpen
  \bibfield  {author} {\bibinfo {author} {\bibfnamefont {A.~R.}\ \bibnamefont {Bell}}\ and\ \bibinfo {author} {\bibfnamefont {J.~G.}\ \bibnamefont {Kirk}},\ }\href {\doibase 10.1103/PhysRevLett.101.200403} {\bibfield  {journal} {\bibinfo  {journal} {Phys. Rev. Lett.}\ }\textbf {\bibinfo {volume} {101}},\ \bibinfo {pages} {200403} (\bibinfo {year} {2008})}\BibitemShut {NoStop}%
\bibitem [{\citenamefont {Fedotov}\ \emph {et~al.}(2010)\citenamefont {Fedotov}, \citenamefont {Narozhny}, \citenamefont {Mourou},\ and\ \citenamefont {Korn}}]{Fedotov:2010ja}%
  \BibitemOpen
  \bibfield  {author} {\bibinfo {author} {\bibfnamefont {A.~M.}\ \bibnamefont {Fedotov}}, \bibinfo {author} {\bibfnamefont {N.~B.}\ \bibnamefont {Narozhny}}, \bibinfo {author} {\bibfnamefont {G.}~\bibnamefont {Mourou}}, \ and\ \bibinfo {author} {\bibfnamefont {G.}~\bibnamefont {Korn}},\ }\href {\doibase 10.1103/PhysRevLett.105.080402} {\bibfield  {journal} {\bibinfo  {journal} {Phys. Rev. Lett.}\ }\textbf {\bibinfo {volume} {105}},\ \bibinfo {pages} {080402} (\bibinfo {year} {2010})},\ \Eprint {http://arxiv.org/abs/1004.5398} {arXiv:1004.5398 [hep-ph]} \BibitemShut {NoStop}%
\bibitem [{\citenamefont {Bulanov}\ \emph {et~al.}(2013)\citenamefont {Bulanov}, \citenamefont {Schroeder}, \citenamefont {Esarey},\ and\ \citenamefont {Leemans}}]{Bulanov:2013cga}%
  \BibitemOpen
  \bibfield  {author} {\bibinfo {author} {\bibfnamefont {S.~S.}\ \bibnamefont {Bulanov}}, \bibinfo {author} {\bibfnamefont {C.~B.}\ \bibnamefont {Schroeder}}, \bibinfo {author} {\bibfnamefont {E.}~\bibnamefont {Esarey}}, \ and\ \bibinfo {author} {\bibfnamefont {W.~P.}\ \bibnamefont {Leemans}},\ }\href {\doibase 10.1103/PhysRevA.87.062110} {\bibfield  {journal} {\bibinfo  {journal} {Phys. Rev. A}\ }\textbf {\bibinfo {volume} {87}},\ \bibinfo {pages} {062110} (\bibinfo {year} {2013})},\ \Eprint {http://arxiv.org/abs/1306.1260} {arXiv:1306.1260 [physics.plasm-ph]} \BibitemShut {NoStop}%
\bibitem [{\citenamefont {Mironov}\ \emph {et~al.}(2021)\citenamefont {Mironov}, \citenamefont {Gelfer},\ and\ \citenamefont {Fedotov}}]{Mironov:2021fft}%
  \BibitemOpen
  \bibfield  {author} {\bibinfo {author} {\bibfnamefont {A.~A.}\ \bibnamefont {Mironov}}, \bibinfo {author} {\bibfnamefont {E.~G.}\ \bibnamefont {Gelfer}}, \ and\ \bibinfo {author} {\bibfnamefont {A.~M.}\ \bibnamefont {Fedotov}},\ }\href {\doibase 10.1103/PhysRevA.104.012221} {\bibfield  {journal} {\bibinfo  {journal} {Phys. Rev. A}\ }\textbf {\bibinfo {volume} {104}},\ \bibinfo {pages} {012221} (\bibinfo {year} {2021})},\ \Eprint {http://arxiv.org/abs/2105.04476} {arXiv:2105.04476 [physics.plasm-ph]} \BibitemShut {NoStop}%
\bibitem [{\citenamefont {Cruz}\ \emph {et~al.}(2022)\citenamefont {Cruz}, \citenamefont {Grismayer}, \citenamefont {Iteanu}, \citenamefont {Tortone},\ and\ \citenamefont {Silva}}]{Cruz:2022zyp}%
  \BibitemOpen
  \bibfield  {author} {\bibinfo {author} {\bibfnamefont {F.}~\bibnamefont {Cruz}}, \bibinfo {author} {\bibfnamefont {T.}~\bibnamefont {Grismayer}}, \bibinfo {author} {\bibfnamefont {S.}~\bibnamefont {Iteanu}}, \bibinfo {author} {\bibfnamefont {P.}~\bibnamefont {Tortone}}, \ and\ \bibinfo {author} {\bibfnamefont {L.~O.}\ \bibnamefont {Silva}},\ }\href {\doibase 10.1063/5.0085847} {\bibfield  {journal} {\bibinfo  {journal} {Phys. Plasmas}\ }\textbf {\bibinfo {volume} {29}},\ \bibinfo {pages} {052902} (\bibinfo {year} {2022})},\ \Eprint {http://arxiv.org/abs/2204.03766} {arXiv:2204.03766 [astro-ph.HE]} \BibitemShut {NoStop}%
\bibitem [{\citenamefont {Mercuri-Baron}\ \emph {et~al.}(2024)\citenamefont {Mercuri-Baron}, \citenamefont {Mironov}, \citenamefont {Riconda}, \citenamefont {Grassi},\ and\ \citenamefont {Grech}}]{Mercuri-Baron:2024rdc}%
  \BibitemOpen
  \bibfield  {author} {\bibinfo {author} {\bibfnamefont {A.}~\bibnamefont {Mercuri-Baron}}, \bibinfo {author} {\bibfnamefont {A.~A.}\ \bibnamefont {Mironov}}, \bibinfo {author} {\bibfnamefont {C.}~\bibnamefont {Riconda}}, \bibinfo {author} {\bibfnamefont {A.}~\bibnamefont {Grassi}}, \ and\ \bibinfo {author} {\bibfnamefont {M.}~\bibnamefont {Grech}},\ }\href@noop {} {\  (\bibinfo {year} {2024})},\ \Eprint {http://arxiv.org/abs/2402.04225} {arXiv:2402.04225 [physics.plasm-ph]} \BibitemShut {NoStop}%
\bibitem [{\citenamefont {Ritus}(1970)}]{Ritus1}%
  \BibitemOpen
  \bibfield  {author} {\bibinfo {author} {\bibfnamefont {V.}~\bibnamefont {Ritus}},\ }\href@noop {} {\bibfield  {journal} {\bibinfo  {journal} {Sov.Phys.JETP}\ ,\ \bibinfo {pages} {1181}} (\bibinfo {year} {1970})}\BibitemShut {NoStop}%
\bibitem [{\citenamefont {Narozhnyi}(1980)}]{Narozhnyi:1980dc}%
  \BibitemOpen
  \bibfield  {author} {\bibinfo {author} {\bibfnamefont {N.~B.}\ \bibnamefont {Narozhnyi}},\ }\href {\doibase 10.1103/PhysRevD.21.1176} {\bibfield  {journal} {\bibinfo  {journal} {Phys. Rev. D}\ }\textbf {\bibinfo {volume} {21}},\ \bibinfo {pages} {1176} (\bibinfo {year} {1980})}\BibitemShut {NoStop}%
\bibitem [{\citenamefont {Fedotov}(2017)}]{Fedotov:2016afw}%
  \BibitemOpen
  \bibfield  {author} {\bibinfo {author} {\bibfnamefont {A.~M.}\ \bibnamefont {Fedotov}},\ }\href {\doibase 10.1088/1742-6596/826/1/012027} {\bibfield  {journal} {\bibinfo  {journal} {J. Phys. Conf. Ser.}\ }\textbf {\bibinfo {volume} {826}},\ \bibinfo {pages} {012027} (\bibinfo {year} {2017})},\ \Eprint {http://arxiv.org/abs/1608.02261} {arXiv:1608.02261 [hep-ph]} \BibitemShut {NoStop}%
\bibitem [{\citenamefont {Luo}\ \emph {et~al.}(2018)\citenamefont {Luo}, \citenamefont {Liu}, \citenamefont {Yuan}, \citenamefont {Chen}, \citenamefont {Yu}, \citenamefont {Li}, \citenamefont {Sorbo}, \citenamefont {Ridgers},\ and\ \citenamefont {Sheng}}]{Luo:2018rkp}%
  \BibitemOpen
  \bibfield  {author} {\bibinfo {author} {\bibfnamefont {W.}~\bibnamefont {Luo}}, \bibinfo {author} {\bibfnamefont {W.-Y.}\ \bibnamefont {Liu}}, \bibinfo {author} {\bibfnamefont {T.}~\bibnamefont {Yuan}}, \bibinfo {author} {\bibfnamefont {M.}~\bibnamefont {Chen}}, \bibinfo {author} {\bibfnamefont {J.-Y.}\ \bibnamefont {Yu}}, \bibinfo {author} {\bibfnamefont {F.-Y.}\ \bibnamefont {Li}}, \bibinfo {author} {\bibfnamefont {D.}~\bibnamefont {Sorbo}}, \bibinfo {author} {\bibfnamefont {C.~P.}\ \bibnamefont {Ridgers}}, \ and\ \bibinfo {author} {\bibfnamefont {Z.-M.}\ \bibnamefont {Sheng}},\ }\href {\doibase 10.1038/s41598-018-26785-8} {\bibfield  {journal} {\bibinfo  {journal} {Sci. Rep.}\ }\textbf {\bibinfo {volume} {8}},\ \bibinfo {pages} {8400} (\bibinfo {year} {2018})}\BibitemShut {NoStop}%
\bibitem [{\citenamefont {Cooper}\ and\ \citenamefont {Mottola}(1989)}]{PhysRevD.40.456}%
  \BibitemOpen
  \bibfield  {author} {\bibinfo {author} {\bibfnamefont {F.}~\bibnamefont {Cooper}}\ and\ \bibinfo {author} {\bibfnamefont {E.}~\bibnamefont {Mottola}},\ }\href {\doibase 10.1103/PhysRevD.40.456} {\bibfield  {journal} {\bibinfo  {journal} {Phys. Rev. D}\ }\textbf {\bibinfo {volume} {40}},\ \bibinfo {pages} {456} (\bibinfo {year} {1989})}\BibitemShut {NoStop}%
\bibitem [{\citenamefont {Gold}\ \emph {et~al.}(2021)\citenamefont {Gold}, \citenamefont {Mcgady}, \citenamefont {Patil},\ and\ \citenamefont {Vardanyan}}]{QED2}%
  \BibitemOpen
  \bibfield  {author} {\bibinfo {author} {\bibfnamefont {G.}~\bibnamefont {Gold}}, \bibinfo {author} {\bibfnamefont {D.~A.}\ \bibnamefont {Mcgady}}, \bibinfo {author} {\bibfnamefont {S.~P.}\ \bibnamefont {Patil}}, \ and\ \bibinfo {author} {\bibfnamefont {V.}~\bibnamefont {Vardanyan}},\ }\href {\doibase 10.1007/JHEP10(2021)072} {\bibfield  {journal} {\bibinfo  {journal} {JHEP}\ }\textbf {\bibinfo {volume} {10}},\ \bibinfo {pages} {072} (\bibinfo {year} {2021})},\ \Eprint {http://arxiv.org/abs/2012.15824} {arXiv:2012.15824 [hep-th]} \BibitemShut {NoStop}%
\bibitem [{\citenamefont {Mihaila}\ \emph {et~al.}(2008)\citenamefont {Mihaila}, \citenamefont {Dawson},\ and\ \citenamefont {Cooper}}]{QED22}%
  \BibitemOpen
  \bibfield  {author} {\bibinfo {author} {\bibfnamefont {B.}~\bibnamefont {Mihaila}}, \bibinfo {author} {\bibfnamefont {J.~F.}\ \bibnamefont {Dawson}}, \ and\ \bibinfo {author} {\bibfnamefont {F.}~\bibnamefont {Cooper}},\ }\href {\doibase 10.1103/PhysRevD.78.116017} {\bibfield  {journal} {\bibinfo  {journal} {Phys. Rev. D}\ }\textbf {\bibinfo {volume} {78}},\ \bibinfo {pages} {116017} (\bibinfo {year} {2008})},\ \Eprint {http://arxiv.org/abs/0811.1353} {arXiv:0811.1353 [hep-ph]} \BibitemShut {NoStop}%
\bibitem [{\citenamefont {Cooper}\ \emph {et~al.}(2008)\citenamefont {Cooper}, \citenamefont {Dawson},\ and\ \citenamefont {Mihaila}}]{QCD}%
  \BibitemOpen
  \bibfield  {author} {\bibinfo {author} {\bibfnamefont {F.}~\bibnamefont {Cooper}}, \bibinfo {author} {\bibfnamefont {J.~F.}\ \bibnamefont {Dawson}}, \ and\ \bibinfo {author} {\bibfnamefont {B.}~\bibnamefont {Mihaila}},\ }in\ \href@noop {} {\emph {\bibinfo {booktitle} {{Conference on Nonequilibrium Phenomena in Cosmology and Particle Physics}}}}\ (\bibinfo {year} {2008})\ \Eprint {http://arxiv.org/abs/0806.1249} {arXiv:0806.1249 [hep-ph]} \BibitemShut {NoStop}%
\bibitem [{\citenamefont {Frantz}(1965)}]{Frantz}%
  \BibitemOpen
  \bibfield  {author} {\bibinfo {author} {\bibfnamefont {L.~M.}\ \bibnamefont {Frantz}},\ }\href@noop {} {\bibfield  {journal} {\bibinfo  {journal} {Phys. Rev.}\ }\textbf {\bibinfo {volume} {139}},\ \bibinfo {pages} {B1326} (\bibinfo {year} {1965})}\BibitemShut {NoStop}%
\bibitem [{\citenamefont {Kibble}(1965)}]{Kibble:1965zza}%
  \BibitemOpen
  \bibfield  {author} {\bibinfo {author} {\bibfnamefont {T.~W.~B.}\ \bibnamefont {Kibble}},\ }\href {\doibase 10.1103/PhysRev.138.B740} {\bibfield  {journal} {\bibinfo  {journal} {Phys. Rev.}\ }\textbf {\bibinfo {volume} {138}},\ \bibinfo {pages} {B740} (\bibinfo {year} {1965})}\BibitemShut {NoStop}%
\bibitem [{\citenamefont {Gavrilov}\ and\ \citenamefont {Gitman}(1990)}]{Gavrilov:1990qa}%
  \BibitemOpen
  \bibfield  {author} {\bibinfo {author} {\bibfnamefont {S.~P.}\ \bibnamefont {Gavrilov}}\ and\ \bibinfo {author} {\bibfnamefont {D.~M.}\ \bibnamefont {Gitman}},\ }\href@noop {} {\bibfield  {journal} {\bibinfo  {journal} {Sov. J. Nucl. Phys.}\ }\textbf {\bibinfo {volume} {51}},\ \bibinfo {pages} {1040} (\bibinfo {year} {1990})}\BibitemShut {NoStop}%
\bibitem [{\citenamefont {Ilderton}\ and\ \citenamefont {Seipt}(2018)}]{Ilderton:2017xbj}%
  \BibitemOpen
  \bibfield  {author} {\bibinfo {author} {\bibfnamefont {A.}~\bibnamefont {Ilderton}}\ and\ \bibinfo {author} {\bibfnamefont {D.}~\bibnamefont {Seipt}},\ }\href {\doibase 10.1103/PhysRevD.97.016007} {\bibfield  {journal} {\bibinfo  {journal} {Phys. Rev. D}\ }\textbf {\bibinfo {volume} {97}},\ \bibinfo {pages} {016007} (\bibinfo {year} {2018})},\ \Eprint {http://arxiv.org/abs/1709.10085} {arXiv:1709.10085 [hep-th]} \BibitemShut {NoStop}%
\bibitem [{\citenamefont {Dvali}\ and\ \citenamefont {Eisemann}(2022)}]{Dvali1}%
  \BibitemOpen
  \bibfield  {author} {\bibinfo {author} {\bibfnamefont {G.}~\bibnamefont {Dvali}}\ and\ \bibinfo {author} {\bibfnamefont {L.}~\bibnamefont {Eisemann}},\ }\href {\doibase 10.1103/PhysRevD.106.125019} {\bibfield  {journal} {\bibinfo  {journal} {Phys. Rev. D}\ }\textbf {\bibinfo {volume} {106}},\ \bibinfo {pages} {125019} (\bibinfo {year} {2022})},\ \Eprint {http://arxiv.org/abs/2211.02618} {arXiv:2211.02618 [hep-th]} \BibitemShut {NoStop}%
\bibitem [{\citenamefont {Dvali}\ \emph {et~al.}(2017)\citenamefont {Dvali}, \citenamefont {Gomez},\ and\ \citenamefont {Zell}}]{Dvali2}%
  \BibitemOpen
  \bibfield  {author} {\bibinfo {author} {\bibfnamefont {G.}~\bibnamefont {Dvali}}, \bibinfo {author} {\bibfnamefont {C.}~\bibnamefont {Gomez}}, \ and\ \bibinfo {author} {\bibfnamefont {S.}~\bibnamefont {Zell}},\ }\href {\doibase 10.1088/1475-7516/2017/06/028} {\bibfield  {journal} {\bibinfo  {journal} {JCAP}\ }\textbf {\bibinfo {volume} {06}},\ \bibinfo {pages} {028} (\bibinfo {year} {2017})},\ \Eprint {http://arxiv.org/abs/1701.08776} {arXiv:1701.08776 [hep-th]} \BibitemShut {NoStop}%
\bibitem [{\citenamefont {Ekman}\ and\ \citenamefont {Ilderton}(2020)}]{Ekman:2020vsc}%
  \BibitemOpen
  \bibfield  {author} {\bibinfo {author} {\bibfnamefont {R.}~\bibnamefont {Ekman}}\ and\ \bibinfo {author} {\bibfnamefont {A.}~\bibnamefont {Ilderton}},\ }\href {\doibase 10.1103/PhysRevD.101.056022} {\bibfield  {journal} {\bibinfo  {journal} {Phys. Rev. D}\ }\textbf {\bibinfo {volume} {101}},\ \bibinfo {pages} {056022} (\bibinfo {year} {2020})},\ \Eprint {http://arxiv.org/abs/2002.03759} {arXiv:2002.03759 [hep-ph]} \BibitemShut {NoStop}%
\bibitem [{\citenamefont {Cristofoli}\ \emph {et~al.}(2022)\citenamefont {Cristofoli}, \citenamefont {Gonzo}, \citenamefont {Kosower},\ and\ \citenamefont {O'Connell}}]{Cristofoli:2021vyo}%
  \BibitemOpen
  \bibfield  {author} {\bibinfo {author} {\bibfnamefont {A.}~\bibnamefont {Cristofoli}}, \bibinfo {author} {\bibfnamefont {R.}~\bibnamefont {Gonzo}}, \bibinfo {author} {\bibfnamefont {D.~A.}\ \bibnamefont {Kosower}}, \ and\ \bibinfo {author} {\bibfnamefont {D.}~\bibnamefont {O'Connell}},\ }\href {\doibase 10.1103/PhysRevD.106.056007} {\bibfield  {journal} {\bibinfo  {journal} {Phys. Rev. D}\ }\textbf {\bibinfo {volume} {106}},\ \bibinfo {pages} {056007} (\bibinfo {year} {2022})},\ \Eprint {http://arxiv.org/abs/2107.10193} {arXiv:2107.10193 [hep-th]} \BibitemShut {NoStop}%
\bibitem [{\citenamefont {Bautista}\ and\ \citenamefont {Siemonsen}(2022)}]{Bautista:2021inx}%
  \BibitemOpen
  \bibfield  {author} {\bibinfo {author} {\bibfnamefont {Y.~F.}\ \bibnamefont {Bautista}}\ and\ \bibinfo {author} {\bibfnamefont {N.}~\bibnamefont {Siemonsen}},\ }\href {\doibase 10.1007/JHEP01(2022)006} {\bibfield  {journal} {\bibinfo  {journal} {JHEP}\ }\textbf {\bibinfo {volume} {01}},\ \bibinfo {pages} {006} (\bibinfo {year} {2022})},\ \Eprint {http://arxiv.org/abs/2110.12537} {arXiv:2110.12537 [hep-th]} \BibitemShut {NoStop}%
\bibitem [{\citenamefont {Adamo}\ \emph {et~al.}(2023)\citenamefont {Adamo}, \citenamefont {Cristofoli}, \citenamefont {Ilderton},\ and\ \citenamefont {Klisch}}]{Adamo:2022qci}%
  \BibitemOpen
  \bibfield  {author} {\bibinfo {author} {\bibfnamefont {T.}~\bibnamefont {Adamo}}, \bibinfo {author} {\bibfnamefont {A.}~\bibnamefont {Cristofoli}}, \bibinfo {author} {\bibfnamefont {A.}~\bibnamefont {Ilderton}}, \ and\ \bibinfo {author} {\bibfnamefont {S.}~\bibnamefont {Klisch}},\ }\href {\doibase 10.1103/PhysRevLett.131.011601} {\bibfield  {journal} {\bibinfo  {journal} {Phys. Rev. Lett.}\ }\textbf {\bibinfo {volume} {131}},\ \bibinfo {pages} {011601} (\bibinfo {year} {2023})},\ \Eprint {http://arxiv.org/abs/2210.04696} {arXiv:2210.04696 [hep-th]} \BibitemShut {NoStop}%
\bibitem [{\citenamefont {Schwinger}(1951)}]{Schwinger:1951nm}%
  \BibitemOpen
  \bibfield  {author} {\bibinfo {author} {\bibfnamefont {J.~S.}\ \bibnamefont {Schwinger}},\ }\href {\doibase 10.1103/PhysRev.82.664} {\bibfield  {journal} {\bibinfo  {journal} {Phys. Rev.}\ }\textbf {\bibinfo {volume} {82}},\ \bibinfo {pages} {664} (\bibinfo {year} {1951})}\BibitemShut {NoStop}%
\bibitem [{\citenamefont {Ilderton}\ and\ \citenamefont {Torgrimsson}(2013)}]{Ilderton:2012qe}%
  \BibitemOpen
  \bibfield  {author} {\bibinfo {author} {\bibfnamefont {A.}~\bibnamefont {Ilderton}}\ and\ \bibinfo {author} {\bibfnamefont {G.}~\bibnamefont {Torgrimsson}},\ }\href {\doibase 10.1103/PhysRevD.87.085040} {\bibfield  {journal} {\bibinfo  {journal} {Phys. Rev. D}\ }\textbf {\bibinfo {volume} {87}},\ \bibinfo {pages} {085040} (\bibinfo {year} {2013})},\ \Eprint {http://arxiv.org/abs/1210.6840} {arXiv:1210.6840 [hep-th]} \BibitemShut {NoStop}%
\bibitem [{\citenamefont {Karbstein}(2019)}]{Karbstein:2019wmj}%
  \BibitemOpen
  \bibfield  {author} {\bibinfo {author} {\bibfnamefont {F.}~\bibnamefont {Karbstein}},\ }\href {\doibase 10.1103/PhysRevLett.122.211602} {\bibfield  {journal} {\bibinfo  {journal} {Phys. Rev. Lett.}\ }\textbf {\bibinfo {volume} {122}},\ \bibinfo {pages} {211602} (\bibinfo {year} {2019})},\ \Eprint {http://arxiv.org/abs/1903.06998} {arXiv:1903.06998 [hep-th]} \BibitemShut {NoStop}%
\bibitem [{\citenamefont {Mironov}\ \emph {et~al.}(2020)\citenamefont {Mironov}, \citenamefont {Meuren},\ and\ \citenamefont {Fedotov}}]{Mironov:2020gbi}%
  \BibitemOpen
  \bibfield  {author} {\bibinfo {author} {\bibfnamefont {A.~A.}\ \bibnamefont {Mironov}}, \bibinfo {author} {\bibfnamefont {S.}~\bibnamefont {Meuren}}, \ and\ \bibinfo {author} {\bibfnamefont {A.~M.}\ \bibnamefont {Fedotov}},\ }\href {\doibase 10.1103/PhysRevD.102.053005} {\bibfield  {journal} {\bibinfo  {journal} {Phys. Rev. D}\ }\textbf {\bibinfo {volume} {102}},\ \bibinfo {pages} {053005} (\bibinfo {year} {2020})},\ \Eprint {http://arxiv.org/abs/2003.06909} {arXiv:2003.06909 [hep-th]} \BibitemShut {NoStop}%
\bibitem [{\citenamefont {Mironov}\ and\ \citenamefont {Fedotov}(2022)}]{Mironov:2021ohk}%
  \BibitemOpen
  \bibfield  {author} {\bibinfo {author} {\bibfnamefont {A.~A.}\ \bibnamefont {Mironov}}\ and\ \bibinfo {author} {\bibfnamefont {A.~M.}\ \bibnamefont {Fedotov}},\ }\href {\doibase 10.1103/PhysRevD.105.033005} {\bibfield  {journal} {\bibinfo  {journal} {Phys. Rev. D}\ }\textbf {\bibinfo {volume} {105}},\ \bibinfo {pages} {033005} (\bibinfo {year} {2022})},\ \Eprint {http://arxiv.org/abs/2109.00634} {arXiv:2109.00634 [hep-th]} \BibitemShut {NoStop}%
\bibitem [{\citenamefont {Edwards}\ and\ \citenamefont {Ilderton}(2021)}]{Edwards:2020npu}%
  \BibitemOpen
  \bibfield  {author} {\bibinfo {author} {\bibfnamefont {J.~P.}\ \bibnamefont {Edwards}}\ and\ \bibinfo {author} {\bibfnamefont {A.}~\bibnamefont {Ilderton}},\ }\href {\doibase 10.1103/PhysRevD.103.016004} {\bibfield  {journal} {\bibinfo  {journal} {Phys. Rev. D}\ }\textbf {\bibinfo {volume} {103}},\ \bibinfo {pages} {016004} (\bibinfo {year} {2021})},\ \Eprint {http://arxiv.org/abs/2010.02085} {arXiv:2010.02085 [hep-ph]} \BibitemShut {NoStop}%
\bibitem [{\citenamefont {Dunne}\ and\ \citenamefont {Harris}(2021)}]{Dunne:2021acr}%
  \BibitemOpen
  \bibfield  {author} {\bibinfo {author} {\bibfnamefont {G.~V.}\ \bibnamefont {Dunne}}\ and\ \bibinfo {author} {\bibfnamefont {Z.}~\bibnamefont {Harris}},\ }\href {\doibase 10.1103/PhysRevD.103.065015} {\bibfield  {journal} {\bibinfo  {journal} {Phys. Rev. D}\ }\textbf {\bibinfo {volume} {103}},\ \bibinfo {pages} {065015} (\bibinfo {year} {2021})},\ \Eprint {http://arxiv.org/abs/2101.10409} {arXiv:2101.10409 [hep-th]} \BibitemShut {NoStop}%
\bibitem [{\citenamefont {Dunne}\ and\ \citenamefont {Harris}(2023)}]{Dunne:2022esi}%
  \BibitemOpen
  \bibfield  {author} {\bibinfo {author} {\bibfnamefont {G.~V.}\ \bibnamefont {Dunne}}\ and\ \bibinfo {author} {\bibfnamefont {Z.}~\bibnamefont {Harris}},\ }\href {\doibase 10.1103/PhysRevD.107.065003} {\bibfield  {journal} {\bibinfo  {journal} {Phys. Rev. D}\ }\textbf {\bibinfo {volume} {107}},\ \bibinfo {pages} {065003} (\bibinfo {year} {2023})},\ \Eprint {http://arxiv.org/abs/2212.04599} {arXiv:2212.04599 [hep-th]} \BibitemShut {NoStop}%
\bibitem [{\citenamefont {Torgrimsson}(2021{\natexlab{a}})}]{Torgrimsson:2021wcj}%
  \BibitemOpen
  \bibfield  {author} {\bibinfo {author} {\bibfnamefont {G.}~\bibnamefont {Torgrimsson}},\ }\href {\doibase 10.1103/PhysRevLett.127.111602} {\bibfield  {journal} {\bibinfo  {journal} {Phys. Rev. Lett.}\ }\textbf {\bibinfo {volume} {127}},\ \bibinfo {pages} {111602} (\bibinfo {year} {2021}{\natexlab{a}})},\ \Eprint {http://arxiv.org/abs/2102.11346} {arXiv:2102.11346 [hep-ph]} \BibitemShut {NoStop}%
\bibitem [{\citenamefont {Torgrimsson}(2021{\natexlab{b}})}]{Torgrimsson:2021zob}%
  \BibitemOpen
  \bibfield  {author} {\bibinfo {author} {\bibfnamefont {G.}~\bibnamefont {Torgrimsson}},\ }\href {\doibase 10.1103/PhysRevD.104.056016} {\bibfield  {journal} {\bibinfo  {journal} {Phys. Rev. D}\ }\textbf {\bibinfo {volume} {104}},\ \bibinfo {pages} {056016} (\bibinfo {year} {2021}{\natexlab{b}})},\ \Eprint {http://arxiv.org/abs/2105.02220} {arXiv:2105.02220 [hep-ph]} \BibitemShut {NoStop}%
\bibitem [{\citenamefont {Torgrimsson}(2023)}]{Torgrimsson:2022ndq}%
  \BibitemOpen
  \bibfield  {author} {\bibinfo {author} {\bibfnamefont {G.}~\bibnamefont {Torgrimsson}},\ }\href {\doibase 10.1103/PhysRevD.107.016019} {\bibfield  {journal} {\bibinfo  {journal} {Phys. Rev. D}\ }\textbf {\bibinfo {volume} {107}},\ \bibinfo {pages} {016019} (\bibinfo {year} {2023})},\ \Eprint {http://arxiv.org/abs/2207.05031} {arXiv:2207.05031 [hep-ph]} \BibitemShut {NoStop}%
\bibitem [{\citenamefont {Podszus}\ and\ \citenamefont {Di~Piazza}(2021)}]{Podszus:2021lms}%
  \BibitemOpen
  \bibfield  {author} {\bibinfo {author} {\bibfnamefont {T.}~\bibnamefont {Podszus}}\ and\ \bibinfo {author} {\bibfnamefont {A.}~\bibnamefont {Di~Piazza}},\ }\href {\doibase 10.1103/PhysRevD.104.016014} {\bibfield  {journal} {\bibinfo  {journal} {Phys. Rev. D}\ }\textbf {\bibinfo {volume} {104}},\ \bibinfo {pages} {016014} (\bibinfo {year} {2021})},\ \Eprint {http://arxiv.org/abs/2103.14637} {arXiv:2103.14637 [hep-ph]} \BibitemShut {NoStop}%
\bibitem [{\citenamefont {Podszus}\ \emph {et~al.}(2022)\citenamefont {Podszus}, \citenamefont {Dinu},\ and\ \citenamefont {Di~Piazza}}]{Podszus:2022jia}%
  \BibitemOpen
  \bibfield  {author} {\bibinfo {author} {\bibfnamefont {T.}~\bibnamefont {Podszus}}, \bibinfo {author} {\bibfnamefont {V.}~\bibnamefont {Dinu}}, \ and\ \bibinfo {author} {\bibfnamefont {A.}~\bibnamefont {Di~Piazza}},\ }\href {\doibase 10.1103/PhysRevD.106.056014} {\bibfield  {journal} {\bibinfo  {journal} {Phys. Rev. D}\ }\textbf {\bibinfo {volume} {106}},\ \bibinfo {pages} {056014} (\bibinfo {year} {2022})},\ \Eprint {http://arxiv.org/abs/2206.10345} {arXiv:2206.10345 [hep-ph]} \BibitemShut {NoStop}%
\bibitem [{\citenamefont {Sudarshan}(1963)}]{sudarshan:1963es}%
  \BibitemOpen
  \bibfield  {author} {\bibinfo {author} {\bibfnamefont {E.~C.~G.}\ \bibnamefont {Sudarshan}},\ }\href {\doibase 10.1103/PhysRevLett.10.277} {\bibfield  {journal} {\bibinfo  {journal} {Phys. Rev. Lett.}\ }\textbf {\bibinfo {volume} {10}},\ \bibinfo {pages} {277} (\bibinfo {year} {1963})}\BibitemShut {NoStop}%
\bibitem [{\citenamefont {Glauber}(1963)}]{glauber:1963ci}%
  \BibitemOpen
  \bibfield  {author} {\bibinfo {author} {\bibfnamefont {R.~J.}\ \bibnamefont {Glauber}},\ }\href {\doibase 10.1103/PhysRev.131.2766} {\bibfield  {journal} {\bibinfo  {journal} {Phys. Rev.}\ }\textbf {\bibinfo {volume} {131}},\ \bibinfo {pages} {2766} (\bibinfo {year} {1963})}\BibitemShut {NoStop}%
\bibitem [{\citenamefont {Gerry}\ and\ \citenamefont {Knight}(2004)}]{GerryKnight2004}%
  \BibitemOpen
  \bibfield  {author} {\bibinfo {author} {\bibfnamefont {C.}~\bibnamefont {Gerry}}\ and\ \bibinfo {author} {\bibfnamefont {P.}~\bibnamefont {Knight}},\ }\href@noop {} {\emph {\bibinfo {title} {Introductory Quantum Optics}}}\ (\bibinfo  {publisher} {Cambridge University Press},\ \bibinfo {year} {2004})\BibitemShut {NoStop}%
\bibitem [{\citenamefont {{Fedotov}}(2009)}]{Fedotov2009exact}%
  \BibitemOpen
  \bibfield  {author} {\bibinfo {author} {\bibfnamefont {A.~M.}\ \bibnamefont {{Fedotov}}},\ }\href {\doibase 10.1134/S1054660X09020108} {\bibfield  {journal} {\bibinfo  {journal} {Laser Physics}\ }\textbf {\bibinfo {volume} {19}},\ \bibinfo {pages} {214} (\bibinfo {year} {2009})}\BibitemShut {NoStop}%
\bibitem [{\citenamefont {Wald}(1975)}]{Wald:1975kc}%
  \BibitemOpen
  \bibfield  {author} {\bibinfo {author} {\bibfnamefont {R.~M.}\ \bibnamefont {Wald}},\ }\href {\doibase 10.1007/BF01609863} {\bibfield  {journal} {\bibinfo  {journal} {Commun. Math. Phys.}\ }\textbf {\bibinfo {volume} {45}},\ \bibinfo {pages} {9} (\bibinfo {year} {1975})}\BibitemShut {NoStop}%
\bibitem [{\citenamefont {Fradkin}\ \emph {et~al.}(1991)\citenamefont {Fradkin}, \citenamefont {Gitman},\ and\ \citenamefont {Shvartsman}}]{Fradkin:1991zq}%
  \BibitemOpen
  \bibfield  {author} {\bibinfo {author} {\bibfnamefont {E.~S.}\ \bibnamefont {Fradkin}}, \bibinfo {author} {\bibfnamefont {D.~M.}\ \bibnamefont {Gitman}}, \ and\ \bibinfo {author} {\bibfnamefont {S.~M.}\ \bibnamefont {Shvartsman}},\ }\href@noop {} {\emph {\bibinfo {title} {{Quantum electrodynamics with unstable vacuum}}}}\ (\bibinfo {year} {1991})\BibitemShut {NoStop}%
\bibitem [{\citenamefont {Gelis}\ and\ \citenamefont {Venugopalan}(2006)}]{Gelis:2006yv}%
  \BibitemOpen
  \bibfield  {author} {\bibinfo {author} {\bibfnamefont {F.}~\bibnamefont {Gelis}}\ and\ \bibinfo {author} {\bibfnamefont {R.}~\bibnamefont {Venugopalan}},\ }\href {\doibase 10.1016/j.nuclphysa.2006.07.020} {\bibfield  {journal} {\bibinfo  {journal} {Nucl. Phys. A}\ }\textbf {\bibinfo {volume} {776}},\ \bibinfo {pages} {135} (\bibinfo {year} {2006})},\ \Eprint {http://arxiv.org/abs/hep-ph/0601209} {arXiv:hep-ph/0601209} \BibitemShut {NoStop}%
\bibitem [{\citenamefont {Gelis}\ and\ \citenamefont {Tanji}(2016)}]{Gelis:2015kya}%
  \BibitemOpen
  \bibfield  {author} {\bibinfo {author} {\bibfnamefont {F.}~\bibnamefont {Gelis}}\ and\ \bibinfo {author} {\bibfnamefont {N.}~\bibnamefont {Tanji}},\ }\href {\doibase 10.1016/j.ppnp.2015.11.001} {\bibfield  {journal} {\bibinfo  {journal} {Prog. Part. Nucl. Phys.}\ }\textbf {\bibinfo {volume} {87}},\ \bibinfo {pages} {1} (\bibinfo {year} {2016})},\ \Eprint {http://arxiv.org/abs/1510.05451} {arXiv:1510.05451 [hep-ph]} \BibitemShut {NoStop}%
\bibitem [{\citenamefont {Kim}(2016)}]{SangPyo}%
  \BibitemOpen
  \bibfield  {author} {\bibinfo {author} {\bibfnamefont {S.~P.}\ \bibnamefont {Kim}},\ }\href@noop {} {\bibfield  {journal} {\bibinfo  {journal} {The Universe}\ }\textbf {\bibinfo {volume} {4}},\ \bibinfo {pages} {8} (\bibinfo {year} {2016})},\ \Eprint {http://arxiv.org/abs/1611.08102} {arXiv:1611.08102 [hep-th]} \BibitemShut {NoStop}%
\bibitem [{\citenamefont {Tomaras}\ \emph {et~al.}(2001)\citenamefont {Tomaras}, \citenamefont {Tsamis},\ and\ \citenamefont {Woodard}}]{Tomaras:2001vs}%
  \BibitemOpen
  \bibfield  {author} {\bibinfo {author} {\bibfnamefont {T.~N.}\ \bibnamefont {Tomaras}}, \bibinfo {author} {\bibfnamefont {N.~C.}\ \bibnamefont {Tsamis}}, \ and\ \bibinfo {author} {\bibfnamefont {R.~P.}\ \bibnamefont {Woodard}},\ }\href {\doibase 10.1088/1126-6708/2001/11/008} {\bibfield  {journal} {\bibinfo  {journal} {JHEP}\ }\textbf {\bibinfo {volume} {11}},\ \bibinfo {pages} {008} (\bibinfo {year} {2001})},\ \Eprint {http://arxiv.org/abs/hep-th/0108090} {arXiv:hep-th/0108090} \BibitemShut {NoStop}%
\bibitem [{\citenamefont {Ilderton}\ and\ \citenamefont {Lindved}(2023)}]{Ilderton:2023ifn}%
  \BibitemOpen
  \bibfield  {author} {\bibinfo {author} {\bibfnamefont {A.}~\bibnamefont {Ilderton}}\ and\ \bibinfo {author} {\bibfnamefont {W.}~\bibnamefont {Lindved}},\ }\href {\doibase 10.1007/JHEP12(2023)118} {\bibfield  {journal} {\bibinfo  {journal} {JHEP}\ }\textbf {\bibinfo {volume} {12}},\ \bibinfo {pages} {118} (\bibinfo {year} {2023})},\ \Eprint {http://arxiv.org/abs/2306.15475} {arXiv:2306.15475 [hep-th]} \BibitemShut {NoStop}%
\bibitem [{\citenamefont {Aoude}\ \emph {et~al.}(2024)\citenamefont {Aoude}, \citenamefont {O'Connell},\ and\ \citenamefont {Sergola}}]{Aoude:2024sve}%
  \BibitemOpen
  \bibfield  {author} {\bibinfo {author} {\bibfnamefont {R.}~\bibnamefont {Aoude}}, \bibinfo {author} {\bibfnamefont {D.}~\bibnamefont {O'Connell}}, \ and\ \bibinfo {author} {\bibfnamefont {M.}~\bibnamefont {Sergola}},\ }\href@noop {} {\  (\bibinfo {year} {2024})},\ \Eprint {http://arxiv.org/abs/2412.05267} {arXiv:2412.05267 [hep-th]} \BibitemShut {NoStop}%
\bibitem [{\citenamefont {Gies}\ \emph {et~al.}(2018)\citenamefont {Gies}, \citenamefont {Karbstein},\ and\ \citenamefont {Kohlf\"urst}}]{Gies:2017ygp}%
  \BibitemOpen
  \bibfield  {author} {\bibinfo {author} {\bibfnamefont {H.}~\bibnamefont {Gies}}, \bibinfo {author} {\bibfnamefont {F.}~\bibnamefont {Karbstein}}, \ and\ \bibinfo {author} {\bibfnamefont {C.}~\bibnamefont {Kohlf\"urst}},\ }\href {\doibase 10.1103/PhysRevD.97.036022} {\bibfield  {journal} {\bibinfo  {journal} {Phys. Rev. D}\ }\textbf {\bibinfo {volume} {97}},\ \bibinfo {pages} {036022} (\bibinfo {year} {2018})},\ \Eprint {http://arxiv.org/abs/1712.03232} {arXiv:1712.03232 [hep-ph]} \BibitemShut {NoStop}%
\bibitem [{\citenamefont {Gies}\ and\ \citenamefont {Karbstein}(2016)}]{GK}%
  \BibitemOpen
  \bibfield  {author} {\bibinfo {author} {\bibfnamefont {H.}~\bibnamefont {Gies}}\ and\ \bibinfo {author} {\bibfnamefont {F.}~\bibnamefont {Karbstein}},\ }\href {\doibase 10.1007/JHEP03(2017)108} {\  (\bibinfo {year} {2016}),\ 10.1007/JHEP03(2017)108},\ \bibinfo {note} {[Addendum: JHEP 03, 108 (2017)]},\ \Eprint {http://arxiv.org/abs/1612.07251} {arXiv:1612.07251 [hep-th]} \BibitemShut {NoStop}%
\bibitem [{\citenamefont {Karbstein}(2017)}]{Karb}%
  \BibitemOpen
  \bibfield  {author} {\bibinfo {author} {\bibfnamefont {F.}~\bibnamefont {Karbstein}},\ }\href {\doibase 10.1007/JHEP10(2017)075} {\bibfield  {journal} {\bibinfo  {journal} {JHEP}\ }\textbf {\bibinfo {volume} {10}},\ \bibinfo {pages} {075} (\bibinfo {year} {2017})},\ \Eprint {http://arxiv.org/abs/1709.03819} {arXiv:1709.03819 [hep-th]} \BibitemShut {NoStop}%
\bibitem [{\citenamefont {Edwards}\ and\ \citenamefont {Schubert}(2017)}]{TadScal}%
  \BibitemOpen
  \bibfield  {author} {\bibinfo {author} {\bibfnamefont {J.~P.}\ \bibnamefont {Edwards}}\ and\ \bibinfo {author} {\bibfnamefont {C.}~\bibnamefont {Schubert}},\ }\href {\doibase 10.1016/j.nuclphysb.2017.08.002} {\bibfield  {journal} {\bibinfo  {journal} {Nucl. Phys. B}\ }\textbf {\bibinfo {volume} {923}},\ \bibinfo {pages} {339} (\bibinfo {year} {2017})},\ \Eprint {http://arxiv.org/abs/1704.00482} {arXiv:1704.00482 [hep-th]} \BibitemShut {NoStop}%
\bibitem [{\citenamefont {Ahmadiniaz}\ \emph {et~al.}(2017)\citenamefont {Ahmadiniaz}, \citenamefont {Bastianelli}, \citenamefont {Corradini}, \citenamefont {Edwards},\ and\ \citenamefont {Schubert}}]{TadSpin}%
  \BibitemOpen
  \bibfield  {author} {\bibinfo {author} {\bibfnamefont {N.}~\bibnamefont {Ahmadiniaz}}, \bibinfo {author} {\bibfnamefont {F.}~\bibnamefont {Bastianelli}}, \bibinfo {author} {\bibfnamefont {O.}~\bibnamefont {Corradini}}, \bibinfo {author} {\bibfnamefont {J.~P.}\ \bibnamefont {Edwards}}, \ and\ \bibinfo {author} {\bibfnamefont {C.}~\bibnamefont {Schubert}},\ }\href {\doibase 10.1016/j.nuclphysb.2017.09.012} {\bibfield  {journal} {\bibinfo  {journal} {Nucl. Phys. B}\ }\textbf {\bibinfo {volume} {924}},\ \bibinfo {pages} {377} (\bibinfo {year} {2017})},\ \Eprint {http://arxiv.org/abs/1704.05040} {arXiv:1704.05040 [hep-th]} \BibitemShut {NoStop}%
\bibitem [{\citenamefont {Ahmadiniaz}\ \emph {et~al.}(2019)\citenamefont {Ahmadiniaz}, \citenamefont {Edwards},\ and\ \citenamefont {Ilderton}}]{Red}%
  \BibitemOpen
  \bibfield  {author} {\bibinfo {author} {\bibfnamefont {N.}~\bibnamefont {Ahmadiniaz}}, \bibinfo {author} {\bibfnamefont {J.~P.}\ \bibnamefont {Edwards}}, \ and\ \bibinfo {author} {\bibfnamefont {A.}~\bibnamefont {Ilderton}},\ }\href {\doibase 10.1007/JHEP05(2019)038} {\bibfield  {journal} {\bibinfo  {journal} {JHEP}\ }\textbf {\bibinfo {volume} {05}},\ \bibinfo {pages} {038} (\bibinfo {year} {2019})},\ \Eprint {http://arxiv.org/abs/1901.09416} {arXiv:1901.09416 [hep-th]} \BibitemShut {NoStop}%
\bibitem [{\citenamefont {Damgaard}\ \emph {et~al.}(2021)\citenamefont {Damgaard}, \citenamefont {Plante},\ and\ \citenamefont {Vanhove}}]{Damgaard:2021ipf}%
  \BibitemOpen
  \bibfield  {author} {\bibinfo {author} {\bibfnamefont {P.~H.}\ \bibnamefont {Damgaard}}, \bibinfo {author} {\bibfnamefont {L.}~\bibnamefont {Plante}}, \ and\ \bibinfo {author} {\bibfnamefont {P.}~\bibnamefont {Vanhove}},\ }\href {\doibase 10.1007/JHEP11(2021)213} {\bibfield  {journal} {\bibinfo  {journal} {JHEP}\ }\textbf {\bibinfo {volume} {11}},\ \bibinfo {pages} {213} (\bibinfo {year} {2021})},\ \Eprint {http://arxiv.org/abs/2107.12891} {arXiv:2107.12891 [hep-th]} \BibitemShut {NoStop}%
\bibitem [{\citenamefont {Damgaard}\ \emph {et~al.}(2023)\citenamefont {Damgaard}, \citenamefont {Hansen}, \citenamefont {Plant\'e},\ and\ \citenamefont {Vanhove}}]{Damgaard:2023ttc}%
  \BibitemOpen
  \bibfield  {author} {\bibinfo {author} {\bibfnamefont {P.~H.}\ \bibnamefont {Damgaard}}, \bibinfo {author} {\bibfnamefont {E.~R.}\ \bibnamefont {Hansen}}, \bibinfo {author} {\bibfnamefont {L.}~\bibnamefont {Plant\'e}}, \ and\ \bibinfo {author} {\bibfnamefont {P.}~\bibnamefont {Vanhove}},\ }\href {\doibase 10.1007/JHEP09(2023)183} {\bibfield  {journal} {\bibinfo  {journal} {JHEP}\ }\textbf {\bibinfo {volume} {09}},\ \bibinfo {pages} {183} (\bibinfo {year} {2023})},\ \Eprint {http://arxiv.org/abs/2307.04746} {arXiv:2307.04746 [hep-th]} \BibitemShut {NoStop}%
\bibitem [{\citenamefont {Gellas}\ \emph {et~al.}(1998)\citenamefont {Gellas}, \citenamefont {Karanikas},\ and\ \citenamefont {Ktorides}}]{PhysRevD.57.3763}%
  \BibitemOpen
  \bibfield  {author} {\bibinfo {author} {\bibfnamefont {G.~C.}\ \bibnamefont {Gellas}}, \bibinfo {author} {\bibfnamefont {A.~I.}\ \bibnamefont {Karanikas}}, \ and\ \bibinfo {author} {\bibfnamefont {C.~N.}\ \bibnamefont {Ktorides}},\ }\href {\doibase 10.1103/PhysRevD.57.3763} {\bibfield  {journal} {\bibinfo  {journal} {Phys. Rev. D}\ }\textbf {\bibinfo {volume} {57}},\ \bibinfo {pages} {3763} (\bibinfo {year} {1998})}\BibitemShut {NoStop}%
\bibitem [{\citenamefont {Laenen}\ \emph {et~al.}(2009)\citenamefont {Laenen}, \citenamefont {Stavenga},\ and\ \citenamefont {White}}]{Laenen2009}%
  \BibitemOpen
  \bibfield  {author} {\bibinfo {author} {\bibfnamefont {E.}~\bibnamefont {Laenen}}, \bibinfo {author} {\bibfnamefont {G.}~\bibnamefont {Stavenga}}, \ and\ \bibinfo {author} {\bibfnamefont {C.~D.}\ \bibnamefont {White}},\ }\href {\doibase 10.1088/1126-6708/2009/03/054} {\bibfield  {journal} {\bibinfo  {journal} {Journal of High Energy Physics}\ }\textbf {\bibinfo {volume} {2009}},\ \bibinfo {pages} {054} (\bibinfo {year} {2009})}\BibitemShut {NoStop}%
\bibitem [{\citenamefont {Koemans~Collado}\ \emph {et~al.}(2018)\citenamefont {Koemans~Collado}, \citenamefont {Di~Vecchia}, \citenamefont {Russo},\ and\ \citenamefont {Thomas}}]{KoemansCollado:2018hss}%
  \BibitemOpen
  \bibfield  {author} {\bibinfo {author} {\bibfnamefont {A.}~\bibnamefont {Koemans~Collado}}, \bibinfo {author} {\bibfnamefont {P.}~\bibnamefont {Di~Vecchia}}, \bibinfo {author} {\bibfnamefont {R.}~\bibnamefont {Russo}}, \ and\ \bibinfo {author} {\bibfnamefont {S.}~\bibnamefont {Thomas}},\ }\href {\doibase 10.1007/JHEP10(2018)038} {\bibfield  {journal} {\bibinfo  {journal} {JHEP}\ }\textbf {\bibinfo {volume} {10}},\ \bibinfo {pages} {038} (\bibinfo {year} {2018})},\ \Eprint {http://arxiv.org/abs/1807.04588} {arXiv:1807.04588 [hep-th]} \BibitemShut {NoStop}%
\bibitem [{\citenamefont {Heissenberg}(2021)}]{Heissenberg:2021tzo}%
  \BibitemOpen
  \bibfield  {author} {\bibinfo {author} {\bibfnamefont {C.}~\bibnamefont {Heissenberg}},\ }\href {\doibase 10.1103/PhysRevD.104.046016} {\bibfield  {journal} {\bibinfo  {journal} {Phys. Rev. D}\ }\textbf {\bibinfo {volume} {104}},\ \bibinfo {pages} {046016} (\bibinfo {year} {2021})},\ \Eprint {http://arxiv.org/abs/2105.04594} {arXiv:2105.04594 [hep-th]} \BibitemShut {NoStop}%
\bibitem [{\citenamefont {Cristofoli}\ \emph {et~al.}(2024)\citenamefont {Cristofoli}, \citenamefont {Gonzo}, \citenamefont {Moynihan}, \citenamefont {O'Connell}, \citenamefont {Ross}, \citenamefont {Sergola},\ and\ \citenamefont {White}}]{Cristofoli:2021jas}%
  \BibitemOpen
  \bibfield  {author} {\bibinfo {author} {\bibfnamefont {A.}~\bibnamefont {Cristofoli}}, \bibinfo {author} {\bibfnamefont {R.}~\bibnamefont {Gonzo}}, \bibinfo {author} {\bibfnamefont {N.}~\bibnamefont {Moynihan}}, \bibinfo {author} {\bibfnamefont {D.}~\bibnamefont {O'Connell}}, \bibinfo {author} {\bibfnamefont {A.}~\bibnamefont {Ross}}, \bibinfo {author} {\bibfnamefont {M.}~\bibnamefont {Sergola}}, \ and\ \bibinfo {author} {\bibfnamefont {C.~D.}\ \bibnamefont {White}},\ }\href {\doibase 10.1007/JHEP06(2024)181} {\bibfield  {journal} {\bibinfo  {journal} {JHEP}\ }\textbf {\bibinfo {volume} {06}},\ \bibinfo {pages} {181} (\bibinfo {year} {2024})},\ \Eprint {http://arxiv.org/abs/2112.07556} {arXiv:2112.07556 [hep-th]} \BibitemShut {NoStop}%
\bibitem [{\citenamefont {Haddad}(2022)}]{Haddad2022}%
  \BibitemOpen
  \bibfield  {author} {\bibinfo {author} {\bibfnamefont {K.}~\bibnamefont {Haddad}},\ }\href {\doibase 10.1103/PhysRevD.105.026004} {\bibfield  {journal} {\bibinfo  {journal} {Phys. Rev. D}\ }\textbf {\bibinfo {volume} {105}},\ \bibinfo {pages} {026004} (\bibinfo {year} {2022})}\BibitemShut {NoStop}%
\bibitem [{\citenamefont {Adamo}\ and\ \citenamefont {Gonzo}(2023)}]{Adamo:2022ooq}%
  \BibitemOpen
  \bibfield  {author} {\bibinfo {author} {\bibfnamefont {T.}~\bibnamefont {Adamo}}\ and\ \bibinfo {author} {\bibfnamefont {R.}~\bibnamefont {Gonzo}},\ }\href {\doibase 10.1007/JHEP05(2023)088} {\bibfield  {journal} {\bibinfo  {journal} {JHEP}\ }\textbf {\bibinfo {volume} {05}},\ \bibinfo {pages} {088} (\bibinfo {year} {2023})},\ \Eprint {http://arxiv.org/abs/2212.13269} {arXiv:2212.13269 [hep-th]} \BibitemShut {NoStop}%
\bibitem [{\citenamefont {Luna}\ \emph {et~al.}(2024)\citenamefont {Luna}, \citenamefont {Moynihan}, \citenamefont {O'Connell},\ and\ \citenamefont {Ross}}]{Luna:2023uwd}%
  \BibitemOpen
  \bibfield  {author} {\bibinfo {author} {\bibfnamefont {A.}~\bibnamefont {Luna}}, \bibinfo {author} {\bibfnamefont {N.}~\bibnamefont {Moynihan}}, \bibinfo {author} {\bibfnamefont {D.}~\bibnamefont {O'Connell}}, \ and\ \bibinfo {author} {\bibfnamefont {A.}~\bibnamefont {Ross}},\ }\href {\doibase 10.1007/JHEP08(2024)045} {\bibfield  {journal} {\bibinfo  {journal} {JHEP}\ }\textbf {\bibinfo {volume} {08}},\ \bibinfo {pages} {045} (\bibinfo {year} {2024})},\ \Eprint {http://arxiv.org/abs/2312.09960} {arXiv:2312.09960 [hep-th]} \BibitemShut {NoStop}%
\bibitem [{\citenamefont {Du}\ \emph {et~al.}(2024)\citenamefont {Du}, \citenamefont {Ajith}, \citenamefont {Rajagopal},\ and\ \citenamefont {Vaman}}]{Yuchen}%
  \BibitemOpen
  \bibfield  {author} {\bibinfo {author} {\bibfnamefont {Y.}~\bibnamefont {Du}}, \bibinfo {author} {\bibfnamefont {S.}~\bibnamefont {Ajith}}, \bibinfo {author} {\bibfnamefont {R.}~\bibnamefont {Rajagopal}}, \ and\ \bibinfo {author} {\bibfnamefont {D.}~\bibnamefont {Vaman}},\ }\href@noop {} {\  (\bibinfo {year} {2024})},\ \Eprint {http://arxiv.org/abs/2409.12895} {arXiv:2409.12895 [hep-th]} \BibitemShut {NoStop}%
\bibitem [{\citenamefont {Di~Vecchia}\ \emph {et~al.}(2024)\citenamefont {Di~Vecchia}, \citenamefont {Heissenberg}, \citenamefont {Russo},\ and\ \citenamefont {Veneziano}}]{DiVecchia:2023frv}%
  \BibitemOpen
  \bibfield  {author} {\bibinfo {author} {\bibfnamefont {P.}~\bibnamefont {Di~Vecchia}}, \bibinfo {author} {\bibfnamefont {C.}~\bibnamefont {Heissenberg}}, \bibinfo {author} {\bibfnamefont {R.}~\bibnamefont {Russo}}, \ and\ \bibinfo {author} {\bibfnamefont {G.}~\bibnamefont {Veneziano}},\ }\href {\doibase 10.1016/j.physrep.2024.06.002} {\bibfield  {journal} {\bibinfo  {journal} {Phys. Rept.}\ }\textbf {\bibinfo {volume} {1083}},\ \bibinfo {pages} {1} (\bibinfo {year} {2024})},\ \Eprint {http://arxiv.org/abs/2306.16488} {arXiv:2306.16488 [hep-th]} \BibitemShut {NoStop}%
\bibitem [{\citenamefont {Ilderton}(2019)}]{Ilderton:2019vot}%
  \BibitemOpen
  \bibfield  {author} {\bibinfo {author} {\bibfnamefont {A.}~\bibnamefont {Ilderton}},\ }\href {\doibase 10.1103/PhysRevD.100.125018} {\bibfield  {journal} {\bibinfo  {journal} {Phys. Rev. D}\ }\textbf {\bibinfo {volume} {100}},\ \bibinfo {pages} {125018} (\bibinfo {year} {2019})},\ \Eprint {http://arxiv.org/abs/1909.02484} {arXiv:1909.02484 [hep-ph]} \BibitemShut {NoStop}%
\bibitem [{\citenamefont {Comtet}(2012)}]{comtet2012advanced}%
  \BibitemOpen
  \bibfield  {author} {\bibinfo {author} {\bibfnamefont {L.}~\bibnamefont {Comtet}},\ }\href {\doibase https://doi.org/10.1007/978-94-010-2196-8} {\emph {\bibinfo {title} {Advanced Combinatorics: The art of finite and infinite expansions}}}\ (\bibinfo  {publisher} {Springer Dordrecht},\ \bibinfo {year} {2012})\BibitemShut {NoStop}%
\bibitem [{\citenamefont {Fukushima}\ \emph {et~al.}(2009)\citenamefont {Fukushima}, \citenamefont {Gelis},\ and\ \citenamefont {Lappi}}]{Fukushima:2009er}%
  \BibitemOpen
  \bibfield  {author} {\bibinfo {author} {\bibfnamefont {K.}~\bibnamefont {Fukushima}}, \bibinfo {author} {\bibfnamefont {F.}~\bibnamefont {Gelis}}, \ and\ \bibinfo {author} {\bibfnamefont {T.}~\bibnamefont {Lappi}},\ }\href {\doibase 10.1016/j.nuclphysa.2009.09.062} {\bibfield  {journal} {\bibinfo  {journal} {Nucl. Phys. A}\ }\textbf {\bibinfo {volume} {831}},\ \bibinfo {pages} {184} (\bibinfo {year} {2009})},\ \Eprint {http://arxiv.org/abs/0907.4793} {arXiv:0907.4793 [hep-ph]} \BibitemShut {NoStop}%
\bibitem [{\citenamefont {Harvey}\ \emph {et~al.}(2015)\citenamefont {Harvey}, \citenamefont {Ilderton},\ and\ \citenamefont {King}}]{Harvey:2014qla}%
  \BibitemOpen
  \bibfield  {author} {\bibinfo {author} {\bibfnamefont {C.~N.}\ \bibnamefont {Harvey}}, \bibinfo {author} {\bibfnamefont {A.}~\bibnamefont {Ilderton}}, \ and\ \bibinfo {author} {\bibfnamefont {B.}~\bibnamefont {King}},\ }\href {\doibase 10.1103/PhysRevA.91.013822} {\bibfield  {journal} {\bibinfo  {journal} {Phys. Rev. A}\ }\textbf {\bibinfo {volume} {91}},\ \bibinfo {pages} {013822} (\bibinfo {year} {2015})},\ \Eprint {http://arxiv.org/abs/1409.6187} {arXiv:1409.6187 [physics.plasm-ph]} \BibitemShut {NoStop}%
\bibitem [{\citenamefont {Khudik}\ \emph {et~al.}(2018)\citenamefont {Khudik} \emph {et~al.}}]{Khudik:2018hkr}%
  \BibitemOpen
  \bibfield  {author} {\bibinfo {author} {\bibfnamefont {V.~N.}\ \bibnamefont {Khudik}} \emph {et~al.},\ }\href {\doibase 10.1063/1.5022640} {\bibfield  {journal} {\bibinfo  {journal} {Phys. Plasmas}\ }\textbf {\bibinfo {volume} {25}},\ \bibinfo {pages} {083104} (\bibinfo {year} {2018})},\ \Eprint {http://arxiv.org/abs/1807.09747} {arXiv:1807.09747 [physics.plasm-ph]} \BibitemShut {NoStop}%
\bibitem [{\citenamefont {Blackburn}\ \emph {et~al.}(2018)\citenamefont {Blackburn}, \citenamefont {Seipt}, \citenamefont {Bulanov},\ and\ \citenamefont {Marklund}}]{Blackburn:2018sfn}%
  \BibitemOpen
  \bibfield  {author} {\bibinfo {author} {\bibfnamefont {T.~G.}\ \bibnamefont {Blackburn}}, \bibinfo {author} {\bibfnamefont {D.}~\bibnamefont {Seipt}}, \bibinfo {author} {\bibfnamefont {S.~S.}\ \bibnamefont {Bulanov}}, \ and\ \bibinfo {author} {\bibfnamefont {M.}~\bibnamefont {Marklund}},\ }\href {\doibase 10.1063/1.5037967} {\bibfield  {journal} {\bibinfo  {journal} {Phys. Plasmas}\ }\textbf {\bibinfo {volume} {25}},\ \bibinfo {pages} {083108} (\bibinfo {year} {2018})},\ \Eprint {http://arxiv.org/abs/1804.11085} {arXiv:1804.11085 [physics.plasm-ph]} \BibitemShut {NoStop}%
\bibitem [{\citenamefont {Blackburn}\ \emph {et~al.}(2023)\citenamefont {Blackburn}, \citenamefont {King},\ and\ \citenamefont {Tang}}]{Blackburn:2023mlo}%
  \BibitemOpen
  \bibfield  {author} {\bibinfo {author} {\bibfnamefont {T.~G.}\ \bibnamefont {Blackburn}}, \bibinfo {author} {\bibfnamefont {B.}~\bibnamefont {King}}, \ and\ \bibinfo {author} {\bibfnamefont {S.}~\bibnamefont {Tang}},\ }\href {\doibase 10.1063/5.0159963} {\bibfield  {journal} {\bibinfo  {journal} {Phys. Plasmas}\ }\textbf {\bibinfo {volume} {30}},\ \bibinfo {pages} {093903} (\bibinfo {year} {2023})},\ \Eprint {http://arxiv.org/abs/2305.13061} {arXiv:2305.13061 [hep-ph]} \BibitemShut {NoStop}%
\bibitem [{\citenamefont {King}\ \emph {et~al.}(2023)\citenamefont {King}, \citenamefont {Heinzl},\ and\ \citenamefont {Blackburn}}]{King:2023eeo}%
  \BibitemOpen
  \bibfield  {author} {\bibinfo {author} {\bibfnamefont {B.}~\bibnamefont {King}}, \bibinfo {author} {\bibfnamefont {T.}~\bibnamefont {Heinzl}}, \ and\ \bibinfo {author} {\bibfnamefont {T.~G.}\ \bibnamefont {Blackburn}},\ }\href {\doibase 10.1140/epjc/s10052-023-12074-w} {\bibfield  {journal} {\bibinfo  {journal} {Eur. Phys. J. C}\ }\textbf {\bibinfo {volume} {83}},\ \bibinfo {pages} {901} (\bibinfo {year} {2023})},\ \Eprint {http://arxiv.org/abs/2307.14417} {arXiv:2307.14417 [hep-ph]} \BibitemShut {NoStop}%
\end{thebibliography}%
\end{document}